\documentclass[conference]{IEEEtran}

\ifCLASSINFOpdf
\else
\fi
\usepackage{graphicx}
\usepackage{xcolor}
\usepackage{caption}

\usepackage{algorithm}
\usepackage{algorithmic}
\usepackage{booktabs}
\usepackage[table]{xcolor}
\usepackage{url}
\usepackage[colorlinks=true, citecolor=blue, linkcolor=blue, urlcolor=blue]{hyperref}
\usepackage{tcolorbox}
\usepackage{amssymb}
\usepackage{placeins}

\usepackage{xcolor}

\usepackage[available,functional,reproduced]{ndssbadges}

\begin{document}
%
\title{Measuring Users' Security and Privacy\\ Concerns About Third-Party\\ Community Bots on Discord}

\author{\IEEEauthorblockN{Kofi Ampomah}
	\IEEEauthorblockA{The University of Alabama\\
		kampomah@crimson.ua.edu}
	\and
	\IEEEauthorblockN{Tu Le}
	\IEEEauthorblockA{The University of Alabama\\
		tle6@ua.edu}
}


%


\IEEEoverridecommandlockouts
\makeatletter\def\@IEEEpubidpullup{6.5\baselineskip}\makeatother
\IEEEpubid{\parbox{\columnwidth}{
        Network and Distributed System Security (NDSS) Symposium 2027\\
        22--26 March 2027, Seoul, Republic of Korea\\
        ISBN 978-1-970672-09-1\\
        https://dx.doi.org/10.14722/ndss.2027.230535\\
        www.ndss-symposium.org
}
\hspace{\columnsep}\makebox[\columnwidth]{}}

\maketitle


\begin{abstract}
Third-party community-based bots are central to how social platforms such as Discord work. They moderate content, manage roles, play music, and support AI-assisted conversations across millions of servers. Yet when a bot is installed in a server, it can access messages, user profiles, and server metadata concerning members who did not participate in the installation decision, with no standardized way to communicate the associated security and privacy implications. In this paper, we analyze 23,102 third-party Discord bots and 183,142 user reviews spanning 49 detected languages and five years of feedback (2021--2026). We employ LLM-based knowledge distillation and fine-tune a DeBERTa-based classifier to identify candidate security and privacy (S\&P)-relevant reviews. After manually validating the complete classifier-positive set, we retain 8,832 reviews identified as S\&P-relevant, representing an observed pipeline yield of 4.82\%, and map them to a 31-theme S\&P taxonomy extending prior work. Our results show that S\&P-related reviews focus heavily on Data Sharing (30.5\%) and Data Collection (16.2\%). Confession and anonymity bots have pipeline-identified SPRRs above 50\%, more than ten times the ecosystem-wide pipeline yield of 4.82\%. Star ratings may obscure the frequency and valence of S\&P-related discussion, while conventional metadata features, such as permission count and administrator status, show only weak associations with user-reported S\&P concerns. We then build BotLens, a proof-of-concept auditing framework that combines review-derived S\&P signals with bot metadata to provide on-demand decision support for the community-based bot ecosystem. BotLens exposes six tools through a Model Context Protocol server, supporting bot-level checkups, server-wide comparison, and integration with AI agents.
\end{abstract}


%
\IEEEpeerreviewmaketitle

\section{Introduction}
\label{sec:introduction}

Social platforms like Discord rely on third-party community bots to moderate content, manage roles, play music, support games, and enable AI-mediated interaction across millions of servers. Discord started as a voice-chat platform for gamers and has grown into a large social platform with more than 200 million monthly active users~\cite{discord2024about}. These bots connect through Discord's API, appear in servers with an APP tag, and provide functionality for entire communities. Unlike mobile apps or browser extensions selected by individual users, a Discord bot is installed server-wide by an administrator through an OAuth2 authorization flow requiring the \texttt{Manage Guild} permission~\cite{discord2024bots}. A single installation decision can therefore expose the data and interactions of potentially thousands of server members to a third-party service over which they have limited control.

Bots may continue accessing server data after installation through Discord's persistent Gateway WebSocket connection, which delivers events such as messages, member joins, reactions, and voice-state changes. A permission bitfield assigned during authorization controls the scope of a bot's access. The Administrator permission grants broad server-level capabilities, while Read Message History allows access to prior messages only in channels the bot is otherwise permitted to view. Privileged Gateway Intents, including the Message Content intent, may further allow bots to access message bodies in available channels and process them in real time. Server members affected by these permissions often have no way to determine what a bot can access or how their data are used. The administrator approving the installation may also misunderstand the scope or consequences of the access granted, consistent with prior findings that users often misinterpret permission scopes and runtime-permission risks~\cite{shen2021permissions}. Discord differs from platforms like Slack, where bots typically operate within private organizational workspaces~\cite{slack2024apps}, and Telegram, which does not expose a comparable structured permission bitfield~\cite{telegram2024bots}. Discord therefore combines large community scale with a formal but difficult-to-observe permission model.

Mobile-app permissions are generally granted by the person who installs and uses the application, while website consent notices are presented directly to visitors during a browsing session~\cite{shen2021permissions,nguyen2022freely}. In contrast, a Discord server admin can install a bot that affects many members who do not participate in the authorization decision. The threat space is broader because overprivileged or compromised bots, rogue admins, and abusive users may exploit bot capabilities within an ongoing social environment. App-store oversight, data-protection requirements, and scrutiny of manipulative consent interfaces provide mechanisms for intervention. Discord members may learn about problematic bot behavior only after installation through incidents, moderation outcomes, or user reviews~\cite{alberts2024computers}. Bot governance is therefore comparatively reactive, increasing the need for clearer, function-specific disclosures at installation time.
Mobile app stores have introduced privacy labels, permission prompts, and review-based transparency mechanisms~\cite{apple2025privacylabels,google2022datasafety}, but community-bot marketplaces provide no comparable standardized disclosure framework. In our dataset, only 0.77\% of bots reference a privacy policy in their listings, leaving users with limited formal information about how bots handle data. User reviews provide an alternative source of evidence because review writers voluntarily describe experiences they considered important enough to report. These reviews offer complementary, unprompted, in-situ evidence of issues that review writers chose to report publicly and, unlike structured studies, are not elicited through researcher-designed questions. However, S\&P concerns in community-bot ecosystems remain underexplored, and no existing tool provides users or administrators with structured, on-demand access to these signals.

\textbf{Research Questions.}
We study the following questions:

\begin{enumerate}
    \item \textbf{RQ1:} What S\&P concerns do Discord bot users raise in reviews, and how are these concerns distributed across bot categories?
    
    \item \textbf{RQ2:} What affective expressions characterize S\&P reviews, and how do these patterns vary across S\&P themes, bot categories and languages?
    
    \item \textbf{RQ3:} How do requested bot permissions and metadata correlate with user-reported S\&P concerns?
\end{enumerate}

\medskip
\textbf{Contributions.}
We make three contributions:

\begin{enumerate}
    \item \textbf{Bot Dataset and S\&P Review Classification:}
    We present a large dataset containing 183,142 user reviews and metadata from 23,102 Discord bots across 49 detected languages and five years of feedback (2021--2026). We develop a 31-theme S\&P taxonomy with 11 themes specific to the bot ecosystem, and we build an automated pipeline for identifying S\&P-relevant reviews. \textbf{(RQ1)}

    \item \textbf{Empirical Findings on Community Bots:}
    We characterize affective-expression patterns in S\&P-relevant review text and examine how they vary across bot categories and S\&P themes, providing a complementary account of how review writers articulate S\&P-related experiences \textbf{(RQ2)}. We also analyze associations between requested permissions, bot metadata, and user-reported S\&P concerns, showing that bot function and data sensitivity are more informative than permission count or administrator status alone \textbf{(RQ3)}.

\item \textbf{BotLens Auditing Framework:}
    We operationalize review-derived S\&P signals through BotLens, a six-tool Model Context Protocol (MCP) server that combines SPRR, sentiment composition, reported S\&P themes, and requested permissions to support on-demand bot-level and server-wide auditing.\footnote{Our artifacts, including the taxonomy, analysis pipeline, and BotLens implementation, are available at \url{https://doi.org/10.5281/zenodo.23583221} and \url{https://github.com/ua-spectral/BotLens}. See~\nameref{appendix:artifact} for details.}

\end{enumerate}

\section{Related Work}
\label{sec:related}

Our work draws on three areas: the security and privacy risks of third-party bots, users' perceptions of bots, and review-based methods for analyzing S\&P concerns at scale.

\subsection{Security and Privacy of Bots on Social Platforms}

Bots and automated agents introduce broader S\&P risks on social platforms~\cite{brundage2018malicious,hendrycks2023overview}. Discord bots frequently request permissions beyond those needed for their stated functionality~\cite{edu2022exploring,yang2023systematic,waheed2022empirical,ye2020chatbot}. Moderation bots mediate rule enforcement and private moderator feedback, while bot adoption depends on moderators' organizational needs and capacities~\cite{seering2024chillbot,kiene2020whousesbots}. Bots in group messaging may also access excessive content and enable cross-context tracking, motivating protocols that preserve end-to-end encryption while limiting bot access~\cite{oesch2020understanding}. SnoopGuard provides selective message access and sender anonymity, but it is preventive and does not assess bots already deployed in existing servers~\cite{chou2025bots}.

Related studies of voice assistants identify unauthorized data access, dormant intents, risky content, and weaknesses in certification and runtime vetting~\cite{guo2020skillexplorer,le2022skillbot,le2024alexa}. Mobile-app research similarly shows that developers struggle with privacy requirements, rely on insecure third-party components, and may retain risky permissions across free and paid applications~\cite{alomar2025effect,alomar2022developers,han2020price}. For messaging-platform chatbots specifically, prior work found that 55\% of chatbots in a leading Discord repository request administrator permission and that only 4.35\% of permission-requesting chatbots provide a privacy policy~\cite{edu2022exploring}. Other work calls for stronger governance of automated systems through Zero Trust principles, continuous monitoring, and improved bot-development practices~\cite{vadlamudi2020rise,siwicki2024protection,vekaria2025big,llamas2025balancing,iliou2021detection,li2021good}. Together, these findings motivate our examination of whether requested permissions and other metadata correspond to the S\&P reviews.

\subsection{User Perceptions of Bots}

The most directly relevant study surveyed 374 participants across five messaging platforms and found that only 41.7\% correctly understood which messages bots could access. Participants reported greater S\&P concern after permissions were disclosed, and the study identified six expectation-mismatch categories and a five-stage S\&P adaptation pathway~\cite{chou2026bot}. Related work examines users' concerns and attitudes toward conversational AI, synthesizes prior research on chatbot privacy, and studies how trust shapes privacy-preserving disclosure behavior~\cite{ali2025understanding,gumusel2025literature,ive2024privacy}.
Bot-generated interactions may increase engagement when they are contextually appropriate, but unchecked automation can also distort platform behavior~\cite{gao2025does,guidi2020users}. Privacy on social platforms is also collective: one user's decisions about shared data can affect others~\cite{squicciarini2010privacy}. This collective dimension is particularly important on Discord, where an administrator can install a bot that affects every member of a server. Our work complements prior surveys and controlled studies by examining S\&P reviews that users voluntarily report.

\subsection{User Review Analysis}

User reviews are an established source for studying S\&P concerns at scale. Prior work has analyzed Android reviews to identify privacy concerns, examine their relationship with app evaluations, and develop systems for navigating privacy feedback~\cite{akgul2024decade,harkous2022hark,nguyen2019short,mukherjee2020empirical,hatamian2019revealing}. Akgul et al.~\cite{akgul2024decade} developed a 20-theme taxonomy for privacy-related Android reviews and showed that reviews provide structured S\&P signals that complement developer disclosures and platform audits. We extend this taxonomy with 11 themes that capture risks specific to community bots.

LLM-generated labels can support large-scale text classification while reducing annotation costs~\cite{rouzegar2024enhancing,pangakis2024knowledge,maekawa2022low}. Building on this approach, we use knowledge distillation with an LLM teacher and a fine-tuned student model to identify S\&P-relevant reviews. Sentiment, topic, and emotion-analysis methods have also been used to extract structured patterns from user-generated text~\cite{naseem2021survey,bordoloi2023sentiment,vora2025hdel}. We combine S\&P classification, thematic analysis, and affective-expression classification to characterize how users describe S\&P-related experiences in the community-bot ecosystem.

\subsection{Our Contribution to Research on Community Bots}

Prior bot research has focused primarily on permission models, protocol-level vulnerabilities, and static or dynamic technical analysis~\cite{edu2022exploring,chou2025bots,waheed2022empirical}. Less is known about how Discord users experience and publicly report S\&P concerns. While prior user studies examine S\&P awareness and attitudes under controlled conditions~\cite{chou2026bot,ive2024privacy}, our work examines unprompted concerns that review writers chose to report publicly in an in-situ platform setting. These reviews provide a distinct source of evidence, although review writers are self-selected and may not represent the broader Discord user population.
Existing review-analysis frameworks were developed primarily for mobile applications and do not capture bot-specific risks such as administrator-mediated installation, gateway intents, and server-wide message access~\cite{akgul2024decade,harkous2022hark,nguyen2019short}. We address this gap with a 31-theme taxonomy adapted to the Discord bot ecosystem and a large-scale analysis of user-reported S\&P concerns. BotLens further operationalizes these review-derived signals through on-demand bot-level and server-wide auditing.
\section{Methodology}
\label{sec:methodology}

\begin{figure*}[ht]
    \includegraphics[width=\textwidth]{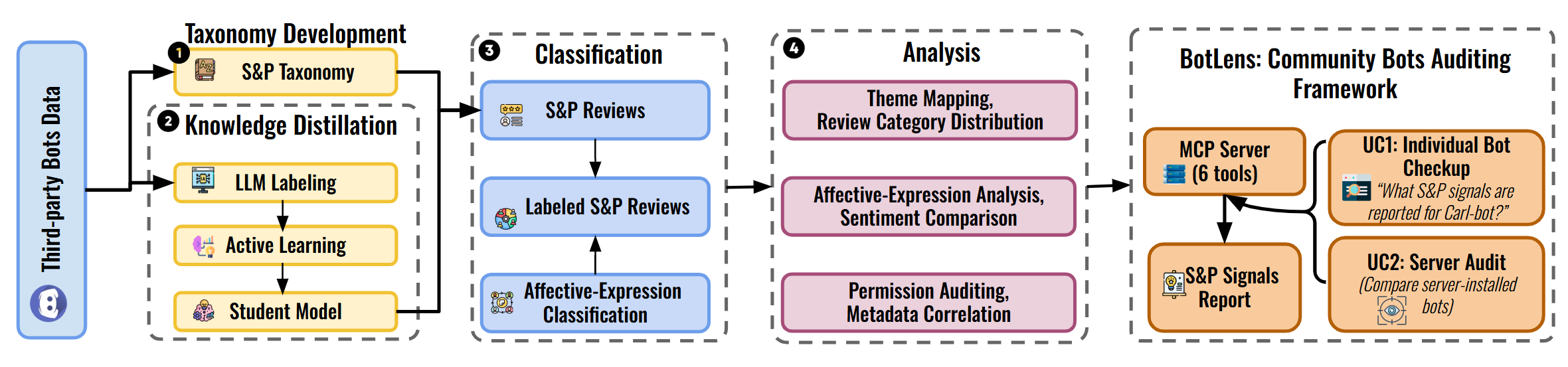}
    \caption{BotLens pipeline: S\&P taxonomy development (Stage 1), classifier training through knowledge distillation (Stage 2), S\&P review identification and affective-expression classification (Stage 3), and review and metadata analysis with BotLens auditing (Stage 4).}
    \label{fig:pipeline}
\end{figure*}

To identify and analyze security and privacy (S\&P) signals in the reviews, we built a four-stage automated pipeline. Stage~1 develops a
Discord-adapted S\&P taxonomy; Stage~2 uses knowledge distillation to train a lightweight classifier from LLM-generated labels; Stage~3 applies that classifier
to the full review corpus and enriches each detected review with taxonomy and affective-expression labels; and Stage~4 uses the resulting dataset to answer our research questions and to power BotLens, an S\&P auditing tool for community-based 	bots. Figure~\ref{fig:pipeline} shows an overview. 

\subsection{Data Collection and Preprocessing}
\label{sec:data}

We built a crawler using Playwright \cite{microsoft_playwright_2026} to collect user reviews and bot metadata from Top.gg, the largest public Discord bot marketplace. For each bot, we collected the full review history alongside bot descriptions, permission flags (encoded as Discord API bitfields), server counts, tags, category assignments, and GitHub repository URLs where available. In total, we collected 201,733 raw reviews spanning five years (January 2021 through February 2026). We preprocessed the raw data in three steps. First, we ran language detection using \texttt{langdetect} and identified 49 distinct languages. Among the 201,733 raw reviews, English accounted for 49.9\% (100,711). Reviews assigned one of the 48 valid non-English language codes accounted for 41.0\% (82,786), while 9.0\% (18,236) had undetermined or invalid language labels. Reviews assigned a valid non-English language code were translated to English using the open-source Argos Translate~\cite{argos_translate_2026}. Table~\ref{app:languages} in Appendix~\ref{app:language_emotions} reports the full language distribution. Second, we applied standard text cleaning: removing reviews shorter than 5 characters, stripping extraneous whitespace, and parsing date fields. Finally, we merged review text with bot metadata and computed derived features, including text length, word count, number of requested permissions, and whether the bot linked a GitHub repository. After preprocessing, 183,142 valid reviews remained. Table~\ref{tab:dataset} summarizes our dataset.

\begin{table}[ht]
\centering
\caption{Dataset overview.}
\label{tab:dataset}
\footnotesize
\begin{tabular}{lr}
\toprule
\textbf{Attribute} & \textbf{Value} \\
\midrule
Total reviews scraped & 201,733 \\
Reviews after preprocessing & 183,142 \\
Unique Discord bots & 23,102 \\
Languages detected & 49 \\
English reviews & 100,711 (49.9\%) \\
Valid non-English reviews & 82,786 (41.0\%) \\
Undetermined/invalid language & 18,236 (9.0\%) \\
Date range & Jan 2021 -- Feb 2026 \\
Source & \url{https://top.gg} \\
\bottomrule
\end{tabular}
\end{table}

\subsection{Stage~1: S\&P Taxonomy Development}
\label{sec:taxonomy}
We developed the taxonomy using a hybrid deductive--inductive thematic analysis~\cite{fereday2006demonstrating,proudfoot2023inductive}, in which an existing coding frame supplies the initial categories and further categories are derived inductively from the data. The deductive component was the 20-theme Android app-review taxonomy of Akgul et al.~\cite{akgul2024decade}, which we applied to our reviews to capture concerns shared with the mobile-app setting. The inductive component addressed concerns that this frame did not cover: two researchers with expertise in security and privacy conducted iterative open coding of 500 S\&P-relevant reviews, assigning codes to concerns that no existing theme accommodated. Discord's developer documentation on Gateway Intents, OAuth2 scopes, and bot permissions served as reference material for interpreting platform-specific terminology in the review text, such as references to intents or permission scopes; it was not used as a source of candidate codes~\cite{discord2024bots}. During codebook development, the researchers discussed candidate codes, resolved disagreements, and iteratively refined the theme definitions. Consensus rather than a reliability coefficient was used at this stage because the researchers were jointly constructing the codebook; agreement statistics for the binary S\&P-relevance labeling are reported in Section~\ref{sec:irr}, where the finalized criteria were applied independently. This process produced 11 new bot-ecosystem themes, including two additional Security \& Safety themes and nine Discord-specific risk themes, resulting in the final 31-theme taxonomy.
The complete taxonomy is presented in Table~\ref{tab:taxonomy} in Appendix~\ref{app:taxonomy}.
Our resulting 31-theme taxonomy includes four categories:
\begin{itemize}
    \item \textbf{Data Practices}: Data Collection, Data Sharing, Data Deletion, Remove PII, Data Exposure, Data Hiding, and Location and Tracking.
    \item \textbf{User Rights \& Controls}: Consent, Privacy Controls, Anonymity/Identification, Advertising, Privacy Policies and Laws, Privacy Invasion, and Positive Privacy.
    \item \textbf{Security \& Safety}: Data Security, Password Issues, Data Accuracy, Safety, Selling Data, Surveillance, Phishing and Scams, and Data Retention.
    \item \textbf{Discord-specific Risks}: Over-Listening, Permission Scope Mismatch, Message Content Access, DM Privacy, Gateway Intent Abuse, Webhook Misuse, Server Profiling, Voice Privacy, and Bot Moderation Transparency.
\end{itemize}

\subsection{Stage~2: Knowledge Distillation}
\label{sec:distillation}

Stage~2 trains a lightweight S\&P classifier through a two-model architecture in which
an LLM acts as the teacher and a fine-tuned smaller model acts as the student. This
design allows us to generate labels for the full 183,142-review corpus without requiring
human annotation at scale, while confining LLM inference to an initial labeling pass.
The S\&P taxonomy from Stage~1 is configured in the teacher system prompt to guide the
classifications.

\subsubsection{LLM for Labeling}
\label{sec:stage2a}

We used Llama~3 8B as the teacher~\cite{llama3}, served locally via Ollama on an NVIDIA
RTX~5090 (32\,GB VRAM). For each review, the LLM produced a JSON output with four
fields: a binary S\&P label, a confidence score (0.0--1.0), a brief reasoning, and the
identified S\&P theme. We set temperature to 0.1 to maximize label consistency and
capped output at 300 tokens per response.

\paragraph{Labeling-budget Comparison}
To examine how the amount of teacher-labeled data affected student-classifier performance, we evaluated five labeling budgets corresponding to 10\%, 20\%, 30\%, 40\%, and 45\% of the 183,142 reviews. At each level, we drew a stratified sample across star ratings (1--5) and text-length bins (short, medium, and long), then trained and evaluated a student classifier using the corresponding teacher-labeled subset. As shown in Table~\ref{tab:labeling_pct}, the 30\% condition achieved the highest observed F1 score among the evaluated configurations (F1\,=\,0.938), with similar precision and recall. Because the labeling conditions used different sampled subsets and evaluation splits, we treat this result as an exploratory comparison rather than evidence that 30\% is a universally optimal labeling budget or that performance saturates beyond this point.

\begin{table}[ht]
\centering
\caption{Student-classifier performance across the evaluated teacher-labeling budgets.}
\label{tab:labeling_pct}

\small
\begin{tabular}{lrrrrr}
\toprule
\textbf{Labeled} & \textbf{Samples} & \textbf{S\&P Reviews} & \textbf{F1} &
\textbf{Prec.} & \textbf{Recall} \\
\midrule
10\% & 18,314 & 69  & 0.727 & 1.000 & 0.571 \\
20\% & 36,628 & 126 & 0.727 & 0.800 & 0.667 \\
\textbf{30\%} & \textbf{54,942} & \textbf{163} & \textbf{0.938} &
\textbf{0.938} & \textbf{0.938} \\
40\% & 73,256 & 216 & 0.833 & 0.741 & 0.952 \\
45\% & 82,413 & 247 & 0.885 & 0.852 & 0.920 \\
\bottomrule
\end{tabular}
\end{table}

To reduce computational cost, we used incremental labeling: the 10\% label set served
as the base, and only the newly selected samples were labeled at each higher percentage.
This cut total labeling time from approximately 200 hours to 54 hours on the RTX~5090.

\subsubsection{Active Learning}
\label{sec:stage2b}

We implemented an active learning loop~\cite{rouzegar2024enhancing} to test whether iterative sample selection could improve on the teacher's labels. The loop uses uncertainty sampling: after training the student on the current labeled set, it selects the unlabeled reviews with prediction probabilities closest to 0.5, which are the examples the current model finds hardest to classify, and sends them to the LLM for labeling in the next round. To prevent the selection from fixating on the decision boundary, we added Gaussian noise ($\varepsilon \sim \mathcal{N}(0, 0.15^2)$) to the uncertainty scores. Each selected sample was labeled three times by the LLM at temperature~0.7, and the majority label was retained to reduce single-pass variability. At the 10\% labeling level, active learning raised F1 from 0.73 to 0.87 with 923 additional labels. At the 30\% labeling level, active learning produced only a small additional improvement, increasing F1 from 0.938 to 0.941. Table~\ref{tab:active_learning} summarizes these results. We interpret this finding only within the evaluated experimental configuration and do not treat it as evidence of a general saturation point.

\begin{table}[t]
\centering
\caption{Active-learning performance at the 10\% and 30\% teacher-labeling levels.}
\label{tab:active_learning}

\small
\setlength{\tabcolsep}{4pt}
\begin{tabular}{lcc}
\toprule
\textbf{Configuration} & \textbf{F1} & \textbf{$\Delta$F1} \\
\midrule
10\% LLM only & 0.727 & --- \\
10\% LLM + AL & 0.870 & +0.143 \\
30\% LLM only & 0.938 & --- \\
30\% LLM + AL & 0.941 & +0.003 \\
\bottomrule
\end{tabular}
\end{table}

\subsubsection{Teacher Model Labeling Validation}
\label{sec:irr}

To assess the reliability of the teacher-generated binary labels, two researchers with expertise in security and privacy independently annotated a stratified 10\% sample of the 54,942-review teacher-labeled subset ($N=5{,}494$). Each researcher independently assigned a binary S\&P-relevance label without access to the teacher-generated label. Pairwise Cohen's~$\kappa$ was 0.74 between the two researchers, 0.71 between the first researcher and Llama~3, and 0.84 between the second researcher and Llama~3. Raw agreement exceeded 99.8\% across all three comparisons. After completing the independent annotations, the researchers discussed and resolved the disagreements. We refer to the adjudicated human annotations as reference labels rather than ground truth.

\subsubsection{Student Model Training}
\label{sec:stage2c}

We fine-tuned DeBERTa-v3-large~\cite{he2021debertav3} (304M parameters) as our student model for binary S\&P classification. Before training, we filtered out LLM labels
with confidence below 0.7 and downsampled the majority class (\texttt{not\_sp}) to a 5:1 ratio relative to the \texttt{sp} class to address label imbalance. We then split
the data into 80\% training, 10\% validation, and 10\% test sets using 	stratified sampling.
We used the following fine-tuning configuration: learning rate $2 \times 10^{-5}$ with 10\% linear warmup, AdamW
optimizer with weight decay 0.01, training batch size 8, evaluation batch size 16, maximum sequence length 512, and FP16 mixed precision. Training ran for up to 10 epochs
with early stopping (patience~3) on validation F1. We set the classification threshold at 0.8 to minimize false positives.

For the classifier trained under the selected 30\% teacher-labeling condition, the corresponding evaluation split contained 98 reviews, including 16 S\&P-relevant and 82 non-S\&P reviews. At the classification threshold of 0.8, the classifier correctly classified 15 of the 16 S\&P-relevant reviews and 81 of the 82 non-S\&P reviews, yielding one false positive and one false negative. Precision, recall, and F1 were each 0.938. The 95\% Wilson confidence intervals for both precision and recall were approximately 0.72--0.99, reflecting substantial uncertainty due to the small number of positive examples. Because these estimates are based on only 16 positive examples, we interpret them as performance under this evaluation setup rather than as precise estimates of generalization performance. Each labeling-budget condition used its own stratified training, validation, and test split, so the differences across Table~\ref{tab:labeling_pct} should be interpreted as exploratory and not as a controlled comparison on a shared test set.

\subsection{Stage~3: Classification}
\label{sec:classification}

With the student model fine-tuned, Stage~3 applies it to the full corpus and enriches each identified S\&P review with S\&P taxonomy assignment and affective-expression classification, producing the structured dataset that feeds Stage~4 analysis.

\subsubsection{S\&P Review Identification}
\label{sec:stage3a}

We applied the trained student classifier to all 183,142 reviews using batch inference (batch size~64, FP16). Reviews with an S\&P relevance score of at least 0.8 were identified as candidates, yielding 8,832 reviews. Both researchers who developed the codebook examined the complete candidate set independently, using the finalized definition of S\&P relevance, and disagreements were resolved through discussion. A review was retained when it discussed a bot-related security or privacy concern, practice, perceived incident, data-handling behavior, or privacy-protective feature. Reviews concerning only functionality, usability, entertainment value, or general dissatisfaction without an S\&P component were outside this definition. The validation assessed whether each review was S\&P-relevant; it did not determine whether the reviewer's underlying claim was factually correct. All 8,832 candidates satisfied the finalized relevance criteria and were retained as the manually validated S\&P-relevant dataset used in the subsequent thematic, sentiment, and affective-expression analyses. The manual validation supports the S\&P-relevance classification of the retained classifier-positive reviews but does not rule out relevant reviews among the classifier-negative set. The teacher-positive rate of \(163/54{,}942=0.30\%\) and the final pipeline yield of \(8{,}832/183{,}142=4.82\%\) arose from different sampling and review procedures using the same taxonomy. Section~\ref{sec:limitations} further discusses the resulting limitations.

\subsubsection{Taxonomy Assignment}
\label{sec:stage3b}

Each manually validated S\&P-relevant review was assigned one primary theme from our 31-theme taxonomy using Llama~3 8B. The prompt supplied the complete taxonomy and its theme definitions and instructed the model to return the most applicable primary theme, a brief description of the issue, and an S\&P sentiment label. We used a temperature of 0.1 to promote consistent outputs. Returned theme labels were programmatically checked against the finalized taxonomy, and outputs that could not be matched were labeled ``unassigned.'' Although secondary themes were retained as descriptive notes during processing, all quantitative analyses use exactly one primary theme per review.

\subsubsection{Affective-Expression Classification}
\label{sec:stage3c}

We analyzed affective expressions in S\&P-relevant review text using the GoEmotions model~\cite{demszky2020goemotions}. The model assigns 28 fine-grained emotion labels, which we aggregated into eight categories: Anger, Disgust, Fear, Sadness, Confusion, Neutral, Joy, and Surprise. We used a multi-label threshold of 0.3. These labels characterize affective language expressed in the reviews and should not be interpreted as direct measurements of users' underlying emotions or longer-term moods. We examine how these textual patterns vary across S\&P themes and bot categories.

\subsection{Stage~4: Data Analysis}
\label{sec:analysis}

Stage~4 performs analyses on the labeled S\&P reviews and bot metadata produced by Stages~1--3, answering RQ1--RQ3 and forming the design of our BotLens auditing framework.
Our analysis relies on two complementary metrics adapted from prior work~\cite{akgul2024decade}. The first metric quantifies the proportion of reviews identified and retained as S\&P-relevant by our pipeline, defined as the Security and Privacy Review Rate (SPRR):

\begin{equation}
\mathrm{SPRR}
=
\frac{N_{\mathrm{validated\ S\&P}}}
{N_{\mathrm{total}}}
\times 100
\end{equation}

$N_{\mathrm{validated\ S\&P}}$ is the number of reviews identified by the classifier and manually validated as S\&P relevant, and $N_{\mathrm{total}}$ is the total number of reviews at the corresponding aggregation level. SPRR is a pipeline-identified review rate. It is not a direct measure of verified technical vulnerabilities, confirmed misconduct, or bot-level risk. SPRR can be scoped to the entire corpus ($SPRR_{all}$), a bot category ($SPRR_{cat}$), a specific tag ($SPRR_{tag}$), an individual bot ($SPRR_{bot}$), or a temporal window ($SPRR_{t}$).

The second metric captures how S\&P concerns distribute across the taxonomy, defined as the Theme Prevalence Rate (TPR):

\begin{equation}
TPR_{\theta} = \frac{N_{\theta}}{N_{sp}} \times 100
\label{eq:tpr}
\end{equation}

\noindent where $N_{\theta}$ is the count of S\&P reviews assigned to theme $\theta$. TPR values across all assigned themes sum to 100\% (excluding reviews labeled ``unassigned''). Together, SPRR identifies \emph{where} pipeline-identified S\&P discussion concentrates across the bot ecosystem, while TPR reveals \emph{what} specific issues users raise within those concentrations.

\paragraph{Temporal trends}
We track $SPRR_{t}$ at 30-day intervals over the five-year observation window (2021--2026). We estimate trend direction through ordinary least-squares regression on the resampled series.

\paragraph{Association tests}
We examine relationships between bot characteristics and SPRR using Pearson~$r$ for continuous linear associations and Spearman~$\rho$ for monotonic relationships in non-normal data.

\paragraph{Permission and transparency analysis} 
Bots request capabilities through a bitfield mechanism encoding 31 distinct flags~\cite{discord_permissions}. We decode these flags into named permissions and examine whether bots that hold high-risk capabilities, such as \textsc{Administrator}, \textsc{Moderate\_Members}, and \textsc{Manage\_Messages}, exhibit elevated SPRR. We also compare SPRR patterns between bots that link a public GitHub repository and those that do not, as a proxy for code transparency.

\section{Findings}
\label{sec:results}

\subsection{Aggregate Security and Privacy Trend Analysis}
\label{sec:aggregate}

We first give an overview of S\&P discussion across the Discord bot ecosystem. Out of the 183,142 reviews, our pipeline identified 8,832 as S\&P-relevant, yielding an observed Security and Privacy Review Rate (SPRR) of 4.82\%.

\begin{figure}[ht]
    \centering
    \includegraphics[width=\columnwidth]{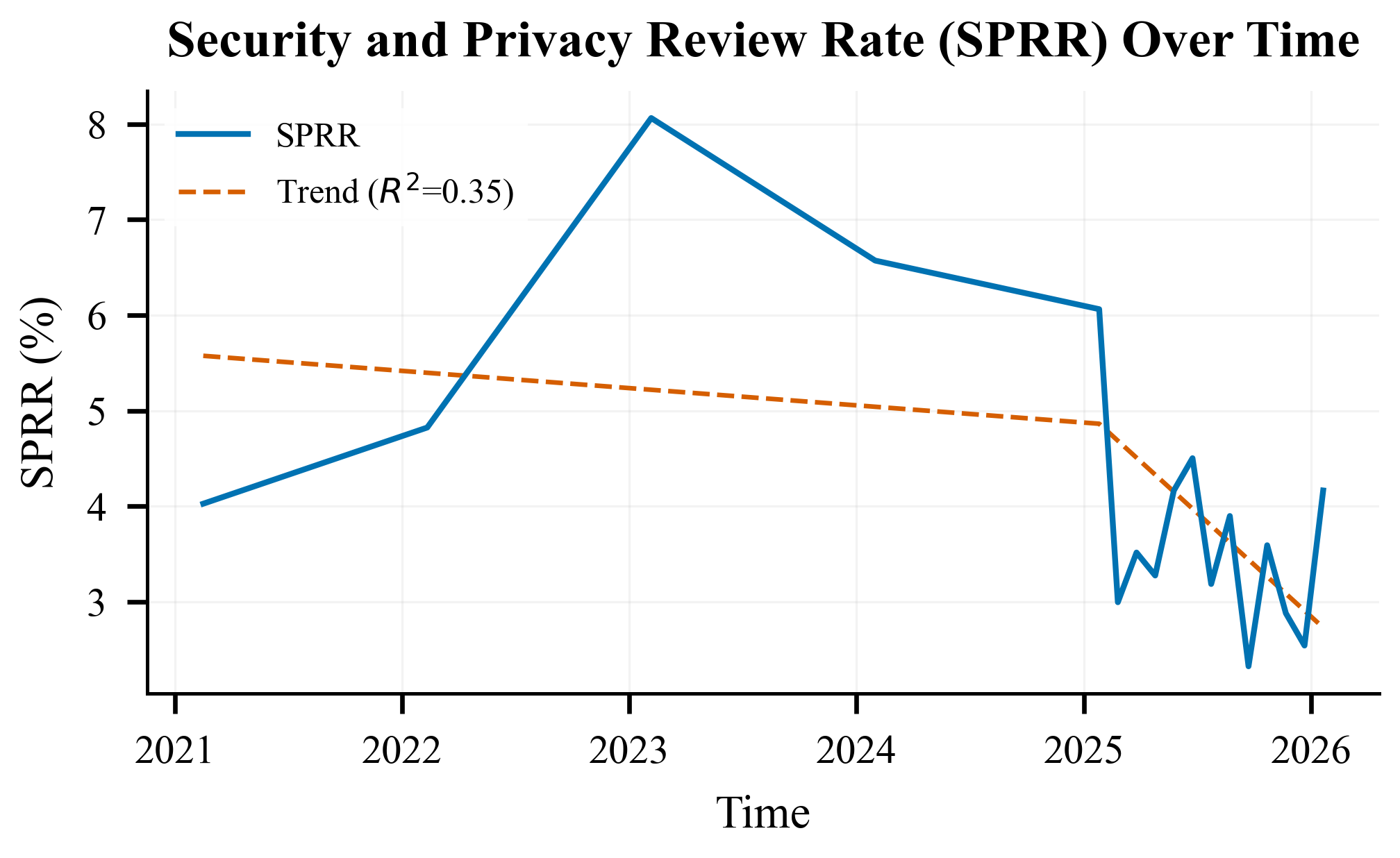}
    \caption{Security and Privacy Review Rate (SPRR) over time in 30-day intervals. SPRR peaked in mid-2023 and has since declined ($R^2 = 0.35$).}
    \label{fig:sprr_over_time}
\end{figure}

Figure~\ref{fig:sprr_over_time} tracks SPRR in 30-day intervals from 2021 through early 2026. The rate climbed from roughly 4\% in early 2021 to a peak around 8\% in mid-2023, then fell to about 3\% by early 2026. A linear regression trend line ($R^2 = 0.35$) indicates a modest downward trend over the full observation window. The rise through 2023 may reflect growing user awareness of S\&P issues during a period of rapid ecosystem expansion. The decline after that might reflect improved bot practices, lower review activity, or shifts in who uses the platform. The increased volatility in 2025 and 2026 lines up with lower review volumes in those periods, which amplifies monthly swings.
This pattern differs from the Android app ecosystem, where prior work observed consistent upward growth in SPRR over a decade (9\% biannual relative growth)~\cite{akgul2024decade}. The Discord ecosystem's five-year window shows a rise-then-decline arc, hinting that S\&P discourse in bot reviews may be more responsive to ecosystem-level changes.

\begin{figure}[ht]
    \centering
    \includegraphics[width=\columnwidth]{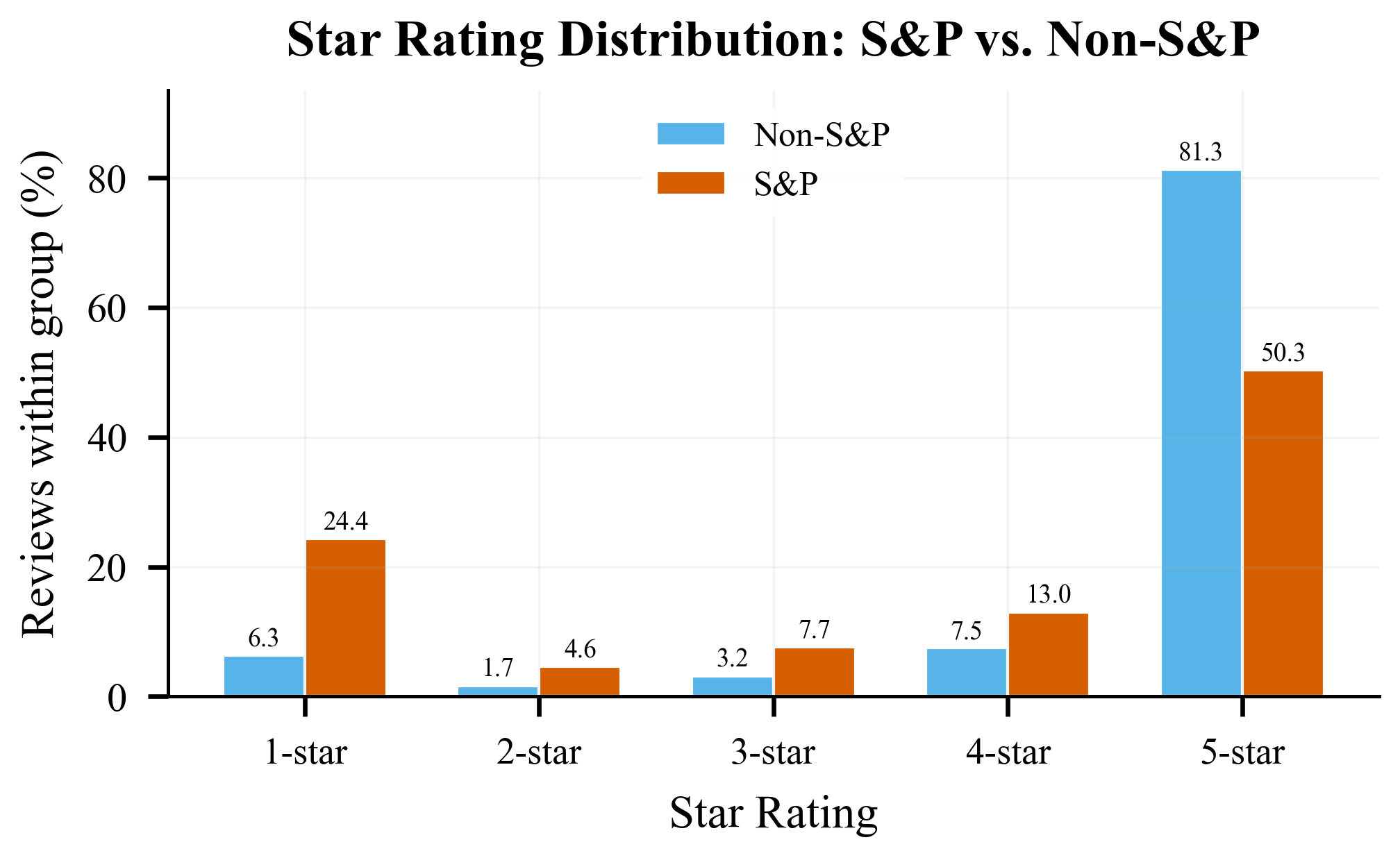}
    \caption{Star-rating distributions for S\&P and non-S\&P reviews.}
    \label{fig:star_rating}
\end{figure}

SPRR does not capture the valence of that discussion. Figure~\ref{fig:star_rating} compares the star-rating distributions of S\&P and non-S\&P reviews. S\&P reviews are more polarized: 50.3\% carry a 5-star rating and 24.4\% carry a 1-star rating, while 4.6\%, 7.7\%, and 13.0\% carry 2-, 3-, and 4-star ratings, respectively. In contrast, non-S\&P reviews are concentrated at 5 stars, which account for 81.3\% of the distribution, followed by 7.5\% at 4 stars, 6.3\% at 1 star, 3.2\% at 3 stars, and 1.7\% at 2 stars. The distributions indicate that S\&P-relevant reviews are more frequently associated with highly positive or highly negative ratings than with mid-range ratings. Highly rated reviews often describe perceived privacy-protective practices, while low-rated reviews report concerns about bot behavior or data practices. These ratings reflect review writers' reported experiences and perceptions rather than independently verified S\&P violations.

\subsection{RQ1: S\&P Themes Across the Bot Ecosystem}
\label{sec:rq1}


We mapped each of the 8,832 S\&P reviews to our 31-theme taxonomy. We start with how concerns spread across the four high-level theme groups, then drill into individual themes.

\subsubsection{Theme Group Distribution}

\begin{figure}[ht]
    \centering
    \includegraphics[width=\columnwidth]{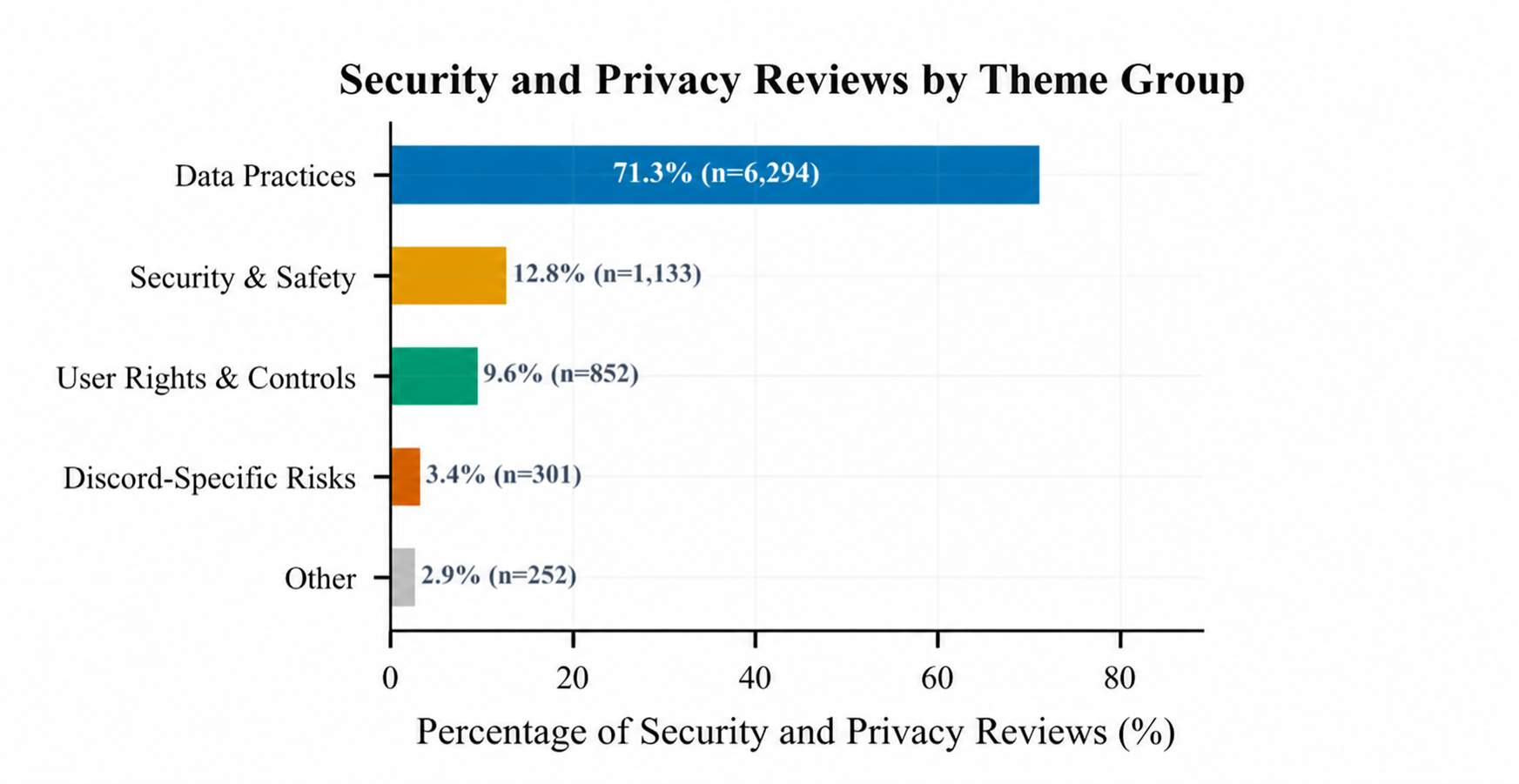}
    \caption{Distribution of security and privacy reviews across the four theme groups. Data Practices accounts for the majority of user concerns.}
    \label{fig:taxonomy_categories}
\end{figure}

Figure~\ref{fig:taxonomy_categories} shows the distribution of S\&P reviews across the four theme groups. Data Practices dominates by a wide margin at 71.3\% of all S\&P reviews (n=6,294). Security \& Safety is a distant second at 12.8\% (n=1,133), then User Rights \& Controls at 9.6\% (n=852), and Discord-Specific Risks at 3.4\% (n=301). The remaining 2.9\% (n=252) were not matched to a specific theme. These results suggest that users prioritize how bots gather, disseminate, and reveal their data over concerns regarding security flaws or platform-specific technical risks.

\subsubsection{Theme-Level Analysis}

\begin{figure}[ht]
    \centering
    \includegraphics[width=\columnwidth]{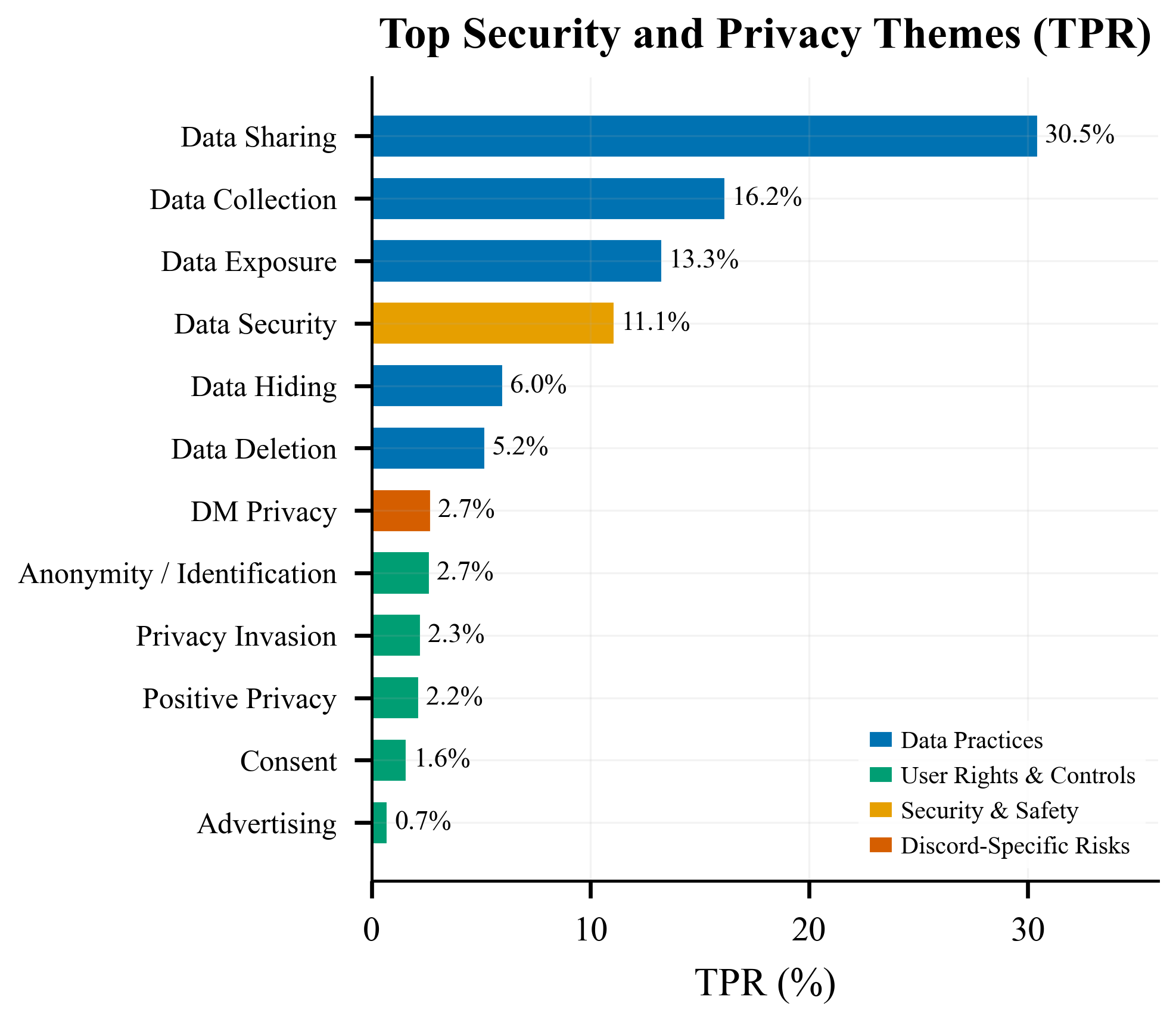}
    \caption{Top S\&P themes by Theme Prevalence Rate}
    \label{fig:theme_distribution}
\end{figure}

Within these categories, Figure~\ref{fig:theme_distribution} shows the Theme Prevalence Rate (TPR) for individual themes. Data Sharing is most prominent at 30.5\%, followed by Data Collection (16.2\%), Data Exposure (13.3\%), and Data Security (11.1\%); together, these four themes account for 71.1\% of S\&P reviews. Data Hiding (6.0\%) and Data Deletion (5.2\%) capture concerns about obscured data practices and missing removal mechanisms. Data Deletion is also the fastest-growing privacy theme reported in the Android ecosystem~\cite{akgul2024decade}. Among Discord-specific themes, DM Privacy is most prominent at 2.7\%. Other themes include Anonymity/Identification (2.7\%), Privacy Invasion (2.3\%), Positive Privacy (2.2\%), Consent (1.6\%), and Advertising (0.7\%). The Positive Privacy theme shows that some users also praise S\&P-protective practices, consistent with prior Android findings~\cite{akgul2024decade}.

\subsubsection{SPRR Across Bot Tags and Categories}

\begin{figure}[!t]
    \centering
    \includegraphics[width=\columnwidth]{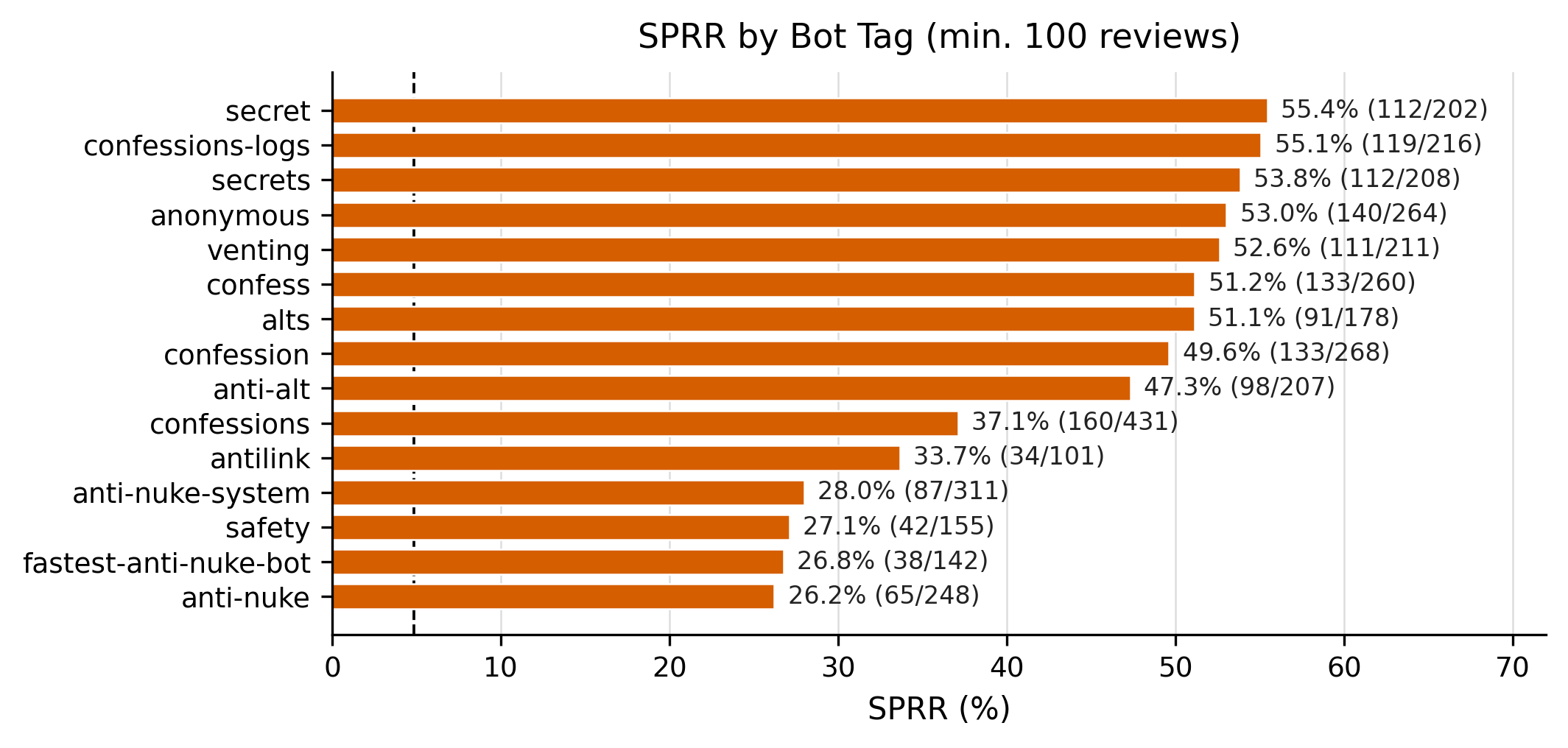}
    \caption{SPRR by bot tag (minimum 100 reviews). Tags related to anonymity and confessions exhibit the highest security and privacy review rates, exceeding 50\%.}
    \label{fig:sprr_by_tag}
\end{figure}

For clarity, several tags in Fig.~\ref{fig:sprr_by_tag} refer to common Discord bot functions. \texttt{secret}, \texttt{secrets}, \texttt{confession}, \texttt{confessions-logs}, and \texttt{anonymous} refer to bots that support anonymous or pseudonymous submissions. \texttt{alts} and \texttt{anti-alt} refer to bots that detect or manage alternate user accounts. \texttt{anti-nuke}, \texttt{anti-nuke-system}, and \texttt{fastest-anti-nuke-bot} refer to security bots intended to prevent destructive administrative actions, while \texttt{antilink} refers to bots that detect or block potentially harmful links.


Figure~\ref{fig:sprr_by_tag} shows the 15 bot tags with the highest SPRR among tags with at least 100 reviews. Anonymity and confession tags dominate: \texttt{secret} has the highest SPRR at 55.4\% (112/202), followed by \texttt{confessions-logs} at 55.1\% (119/216), \texttt{secrets} at 53.8\% (112/208), and \texttt{anonymous} at 53.0\% (140/264). \texttt{venting} (52.6\%), \texttt{confess} (51.2\%), and \texttt{alts} (51.1\%) also exceed 50\%. For bots carrying these labels, our pipeline identified more than half of their reviews as S\&P-related. These rates exceed the ecosystem-wide pipeline yield by more than ten times. These bots handle sensitive disclosures, likely increasing scrutiny of their data practices. The high SPRR for \texttt{alts} and \texttt{anti-alt} (51.1\% and 47.3\%) similarly reflects their handling of identity information. Security-oriented tags form a second cluster: \texttt{antilink} (33.7\%), \texttt{anti-nuke-system} (28.0\%), \texttt{safety} (27.1\%), \texttt{fastest-anti-nuke-bot} (26.8\%), and \texttt{anti-nuke} (26.2\%). These bots require elevated permissions for protective functions, creating a tension between security functionality and broad access.

\subsubsection{Theme Variation Across Bot Categories}

\begin{figure}[ht]
    \centering
    \includegraphics[width=\columnwidth]{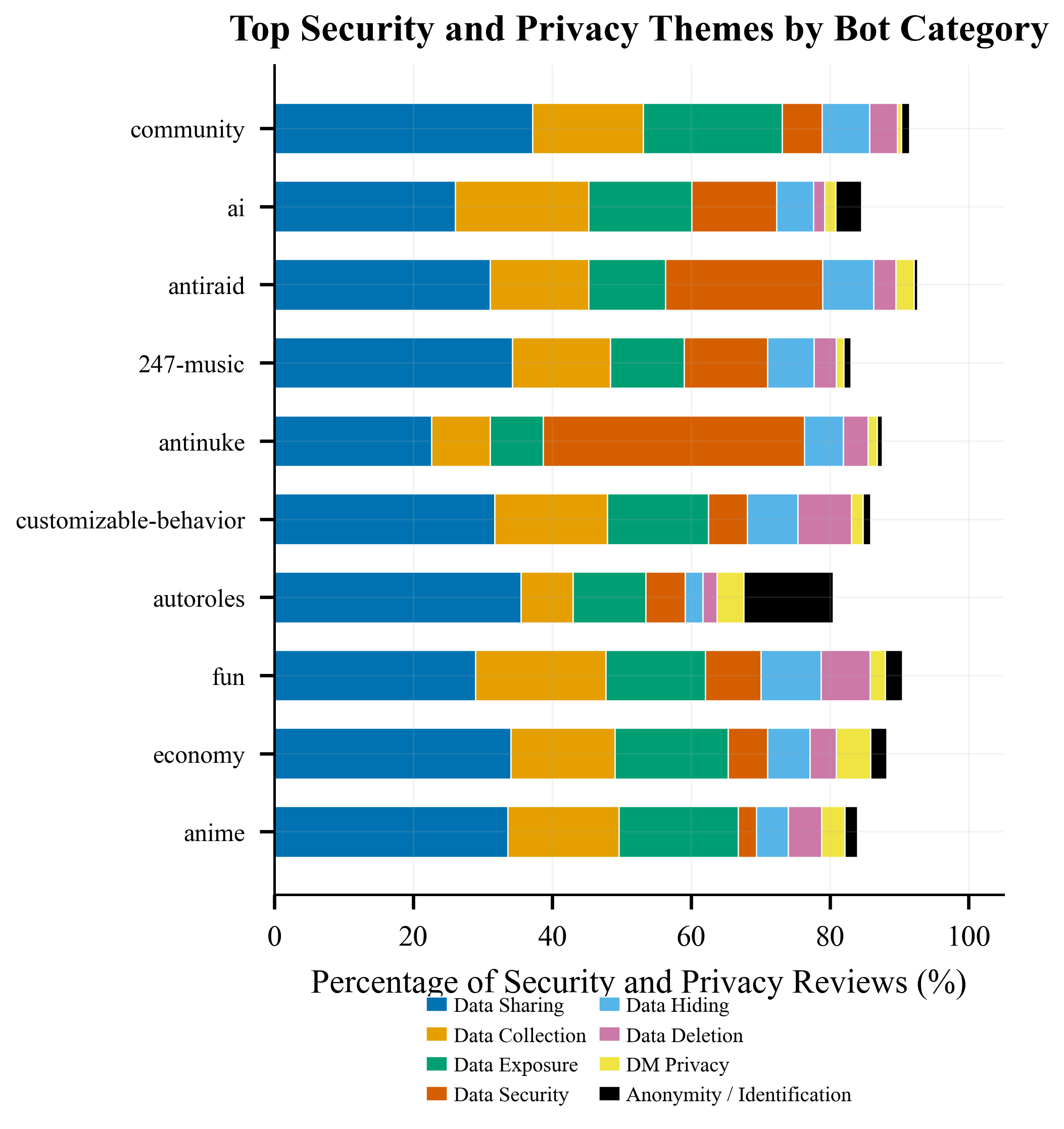}
    \caption{Top S\&P themes by bot category. While Data Sharing dominates most bot categories, the relative contribution of other themes varies by bot function}
    \label{fig:themes_by_category}
\end{figure}

Data Sharing dominates at 30.5\%, but theme composition varies by bot function. Figure~\ref{fig:themes_by_category} shows the theme composition for the top 10 categories by S\&P review volume. Data Sharing is especially prominent among community and economy bots ($\sim$35\%). AI bots show larger shares of Data Collection and Data Exposure, while antinuke and antiraid bots show substantially more Data Security discussion. The 247-music category has elevated Data Collection, and autoroles bots show more Anonymity/Identification concerns. These differences indicate that S\&P discussion varies with bot functionality and data sensitivity.

\subsection{RQ2: Affective Expressions in S\&P Reviews}
\label{sec:rq2}

RQ1 established what users report and where those reports occur. RQ2 examines how affective language is expressed in the review text. We analyze affective-expression patterns across S\&P themes and bot categories and relate these patterns to the star-rating polarization reported in Section~\ref{sec:aggregate}.

\subsubsection{Affective Expressions by S\&P Theme}

\begin{figure}[ht]
    \centering
    \includegraphics[width=\columnwidth]{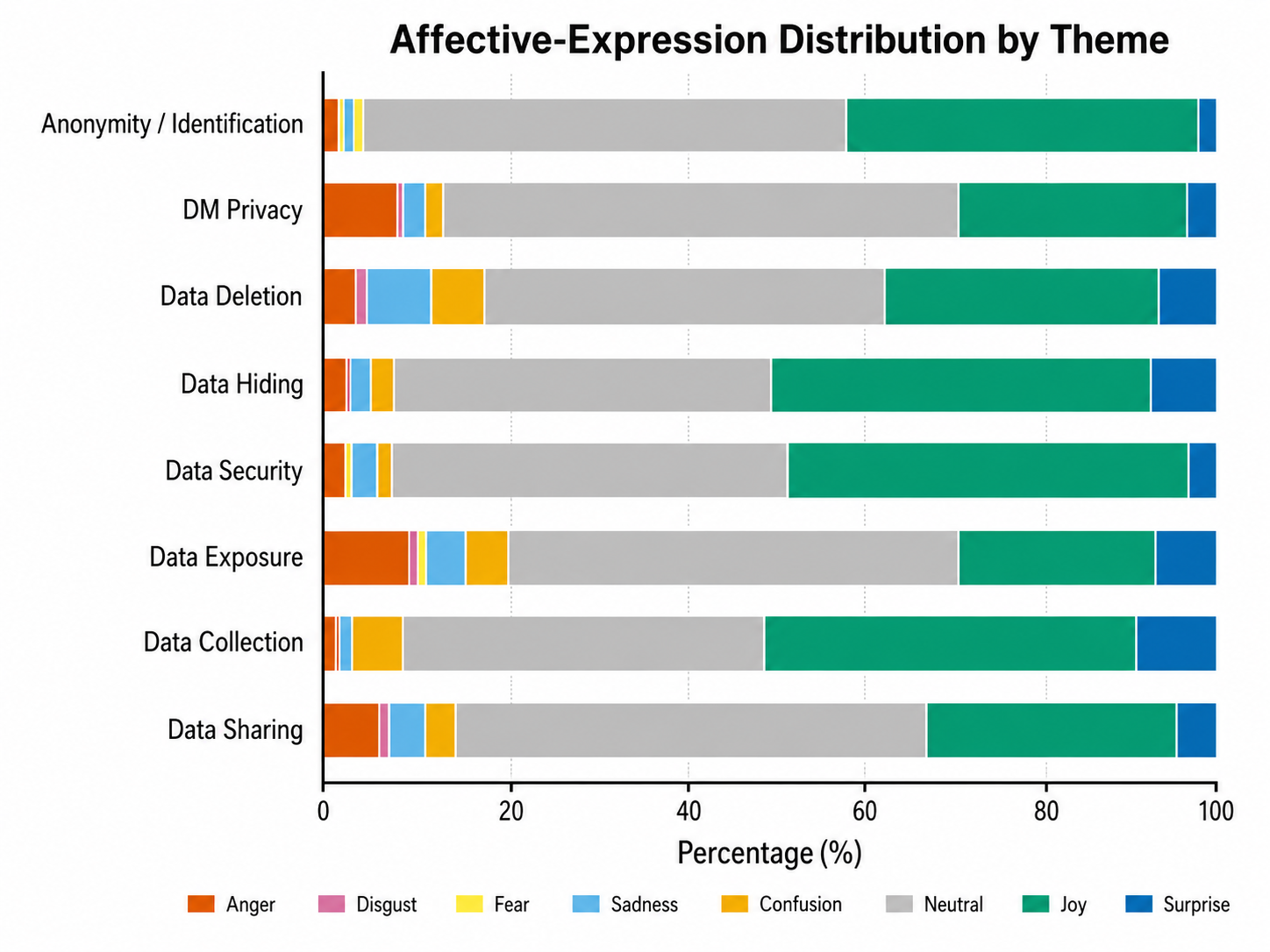}
    \caption{Distribution of affective-expression labels across the top S\&P themes. Neutral is the most frequent label across themes. Data Exposure has the highest proportions of Anger and Confusion, while DM Privacy also contains a notable proportion of Anger labels.}
    \label{fig:emotions_by_theme}
\end{figure}

Figure~\ref{fig:emotions_by_theme} shows affective-expression labels across the top eight S\&P themes. Neutral dominates all themes at roughly 40\% to 60\%, indicating that many S\&P reviews use matter-of-fact language. Joy is also common, consistent with reviews praising privacy-protective behavior. Data Exposure has the largest shares of Anger ($\sim$10\%) and Confusion ($\sim$8\%), while DM Privacy also shows notable Anger. Data Deletion has comparatively more Joy, consistent with praise for removal mechanisms, and Data Security shows more Surprise in reviews describing unexpected security incidents.

\subsubsection{Affective Expressions by Bot Category}

\begin{figure}[ht]
    \centering
    \includegraphics[width=\columnwidth]{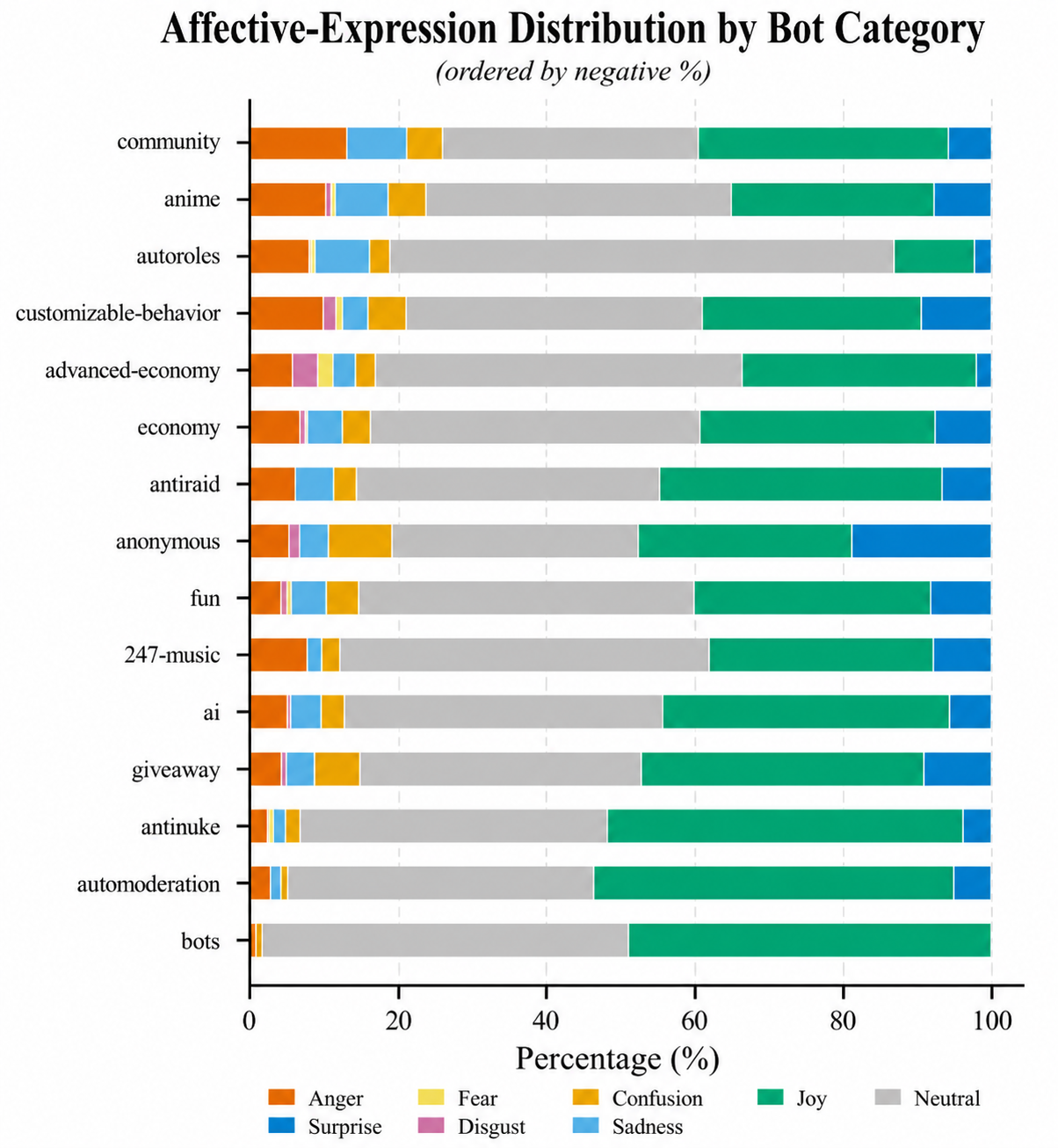}
    \caption{Distribution of affective-expression labels across bot categories, ordered by the combined proportion of negative labels. Community bots have the highest combined proportion of Anger, Sadness, Fear, Disgust, and Confusion labels.}
    \label{fig:emotions_by_category}
\end{figure}

Figure~\ref{fig:emotions_by_category} presents the distribution of affective-expression labels across bot categories, ordered by the proportion of negative labels. Community bots have the highest proportion of negative affective-expression labels, with Anger accounting for roughly 12\% of their S\&P reviews, followed by Sadness and Confusion. Anime and autoroles bots also show comparatively elevated proportions of negative labels. These categories commonly support social interactions, identity-related features, and community management, which may help explain why reviews discussing S\&P concerns use more negative affective language.

Automoderation and antinuke bots have the lowest proportions of negative labels and comparatively high proportions of Joy and Neutral. Reviews of these bots often describe protective or administrative functions, and positive affective language appears when those functions are perceived as effective. Giveaway bots contain a notable proportion of Confusion labels, which may reflect uncertainty expressed in reviews about how participant data are collected or used. AI bots show a more mixed distribution, with moderate Anger and a comparatively large Joy component, indicating that S\&P concerns and positive evaluations of functionality can appear within the same bot category.

\subsubsection{Connecting Affective Expressions to Star Ratings}


Affective-expression labels broadly align with the star-rating polarization in Figure~\ref{fig:star_rating}. Neutral and Joy are common, while Anger and Confusion are more prominent for Data Exposure and DM Privacy. These patterns characterize textual affect rather than users' underlying emotions or decision-making processes. Appendix~\ref{app:language_emotions} provides a breakdown by detected language.

\subsection{RQ3: Permissions, Metadata, and S\&P Concerns}
\label{sec:rq3}


RQ1 showed that SPRR varies substantially with bot function, particularly with the type of content a bot handles. A follow-up question is whether requested permissions and other metadata predict where S\&P discussion concentrates. This section explores this aspect across several metadata features: permission count, specific high-risk permissions, server count, administrator status, and code transparency. The analyses in this section operate at the bot level and are restricted to the 1,457 bots with at least five reviews, of which at least one was classified as S\&P-relevant. The server-count analysis is further restricted to the 896 of these bots that publicly disclose a server count on their listing.

\subsubsection{Permission Count and SPRR}

\begin{figure}[ht]
    \centering
    \includegraphics[width=\columnwidth]{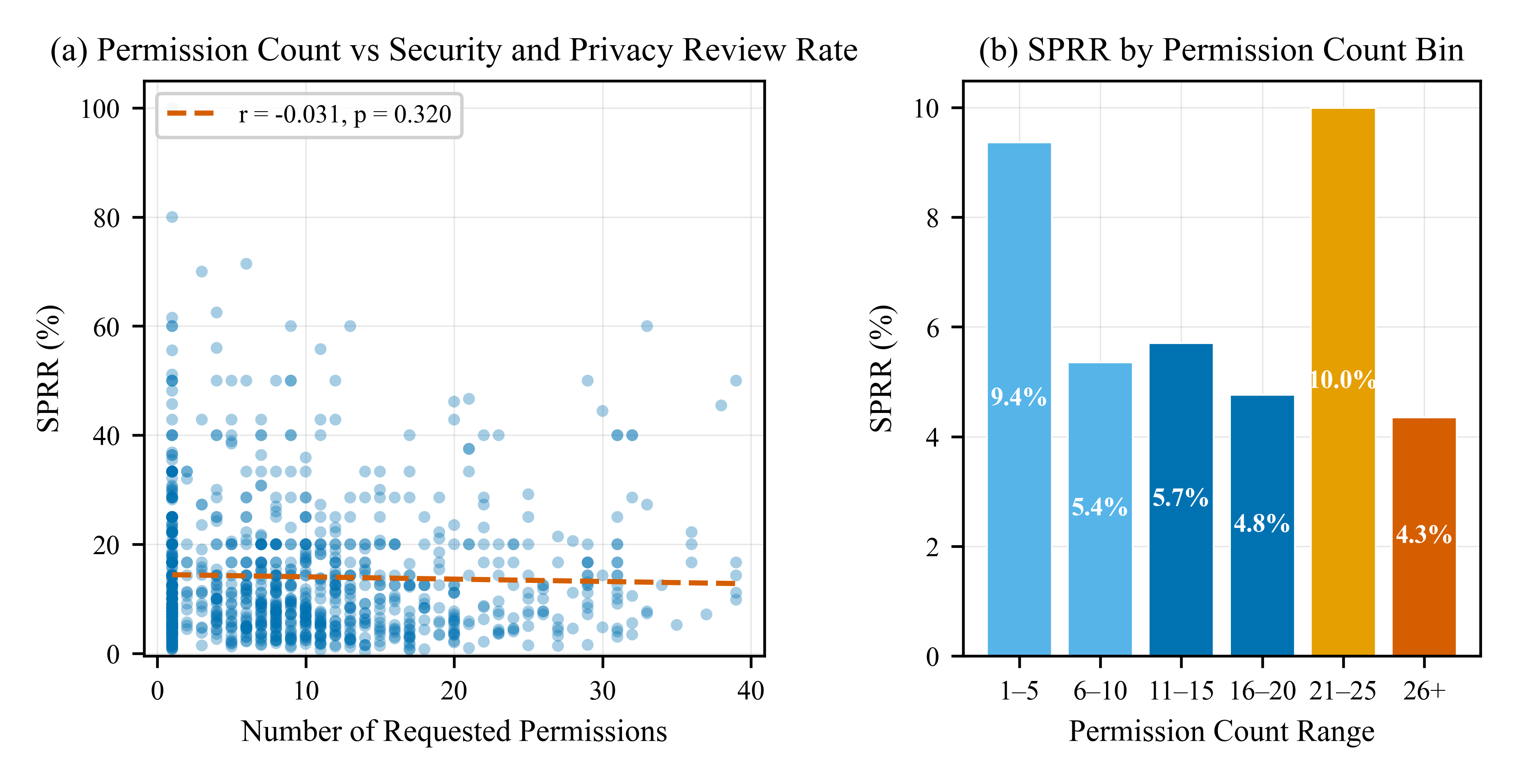}
    \caption{(a) Scatter plot of permission count vs.\ SPRR showing no significant linear correlation ($r = -0.031$, $p = 0.320$). (b) SPRR by permission count bin reveals a non-monotonic pattern.}
    \label{fig:rq4_perm_count}
\end{figure}

Figure~\ref{fig:rq4_perm_count}(a) shows no significant linear relationship between requested-permission count and SPRR ($r=-0.031$, $p=0.320$), with substantial variance across permission levels. Figure~\ref{fig:rq4_perm_count}(b) shows a non-monotonic pattern: SPRR is 9.4\% for 1--5 permissions, 5.4\% for 6--10, 5.7\% for 11--15, 4.8\% for 16--20, 10.0\% for 21--25, and 4.3\% for 26+. Thus, permission count alone poorly predicts S\&P discussion. The elevated SPRR among low-permission bots partly reflects confession and anonymity bots from RQ1, which request limited access while handling sensitive content.

\subsubsection{High-Risk Permissions}

\begin{figure}[ht]
    \centering
    \includegraphics[width=\columnwidth]{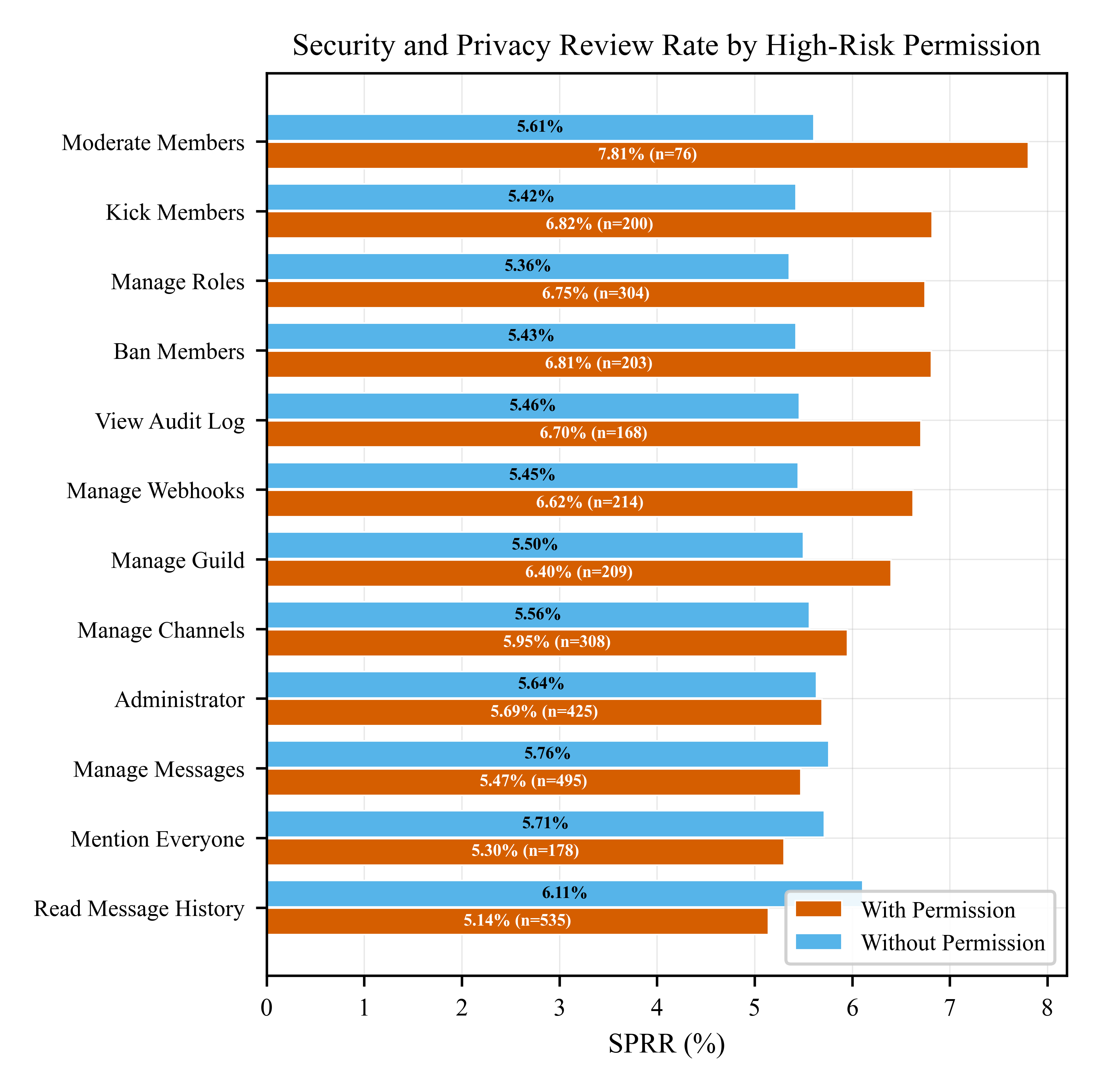}
    \caption{SPRR comparison for bots with and without each high-risk permission. Nine of the twelve permissions show higher SPRR when requested, while three show the opposite pattern.}
    \label{fig:rq4_high_risk}
\end{figure}

Figure~\ref{fig:rq4_high_risk} compares SPRR for bots with and without 12 high-risk permissions. Differences are generally modest. The largest gaps occur for \textsc{Moderate Members} (7.81\% vs.\ 5.61\%, n=76), \textsc{Kick Members} (6.82\% vs.\ 5.42\%, n=200), and \textsc{Ban Members} (6.81\% vs.\ 5.43\%, n=203). \textsc{Administrator} shows almost no difference: 5.69\% (n=425) versus 5.64\% (n=1,032). No individual high-risk permission shows a strong association with S\&P discussion, motivating a closer comparison of admin and non-admin bots.

\subsubsection{Admin vs.\ Non-Admin Bots}

\begin{figure}[ht]
    \centering
    \includegraphics[width=\columnwidth]{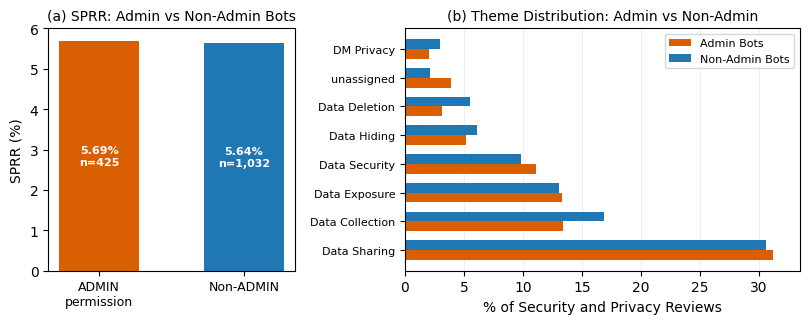}
    \caption{(a) SPRR for admin vs.\ non-admin bots shows virtually identical rates (5.69\% vs.\ 5.64\%). (b) Theme distributions are also nearly identical between the two groups.}
    \label{fig:rq4_admin}
\end{figure}

Figure~\ref{fig:rq4_admin}(a) shows nearly identical SPRR for admin and non-admin bots, 5.69\% (n=425) versus 5.64\% (n=1,032). Figure~\ref{fig:rq4_admin}(b) likewise shows similar theme profiles, dominated by Data Sharing, Data Collection, Data Exposure, and Data Security. Non-admin bots show slightly more Data Collection and DM Privacy, while admin bots show marginally more Data Sharing. These results suggest that S\&P discussion is associated more with bot function and category (RQ1) than with administrator permission.

\subsubsection{Server Count and SPRR}

\begin{figure}[ht]
    \centering
    \includegraphics[width=\columnwidth]{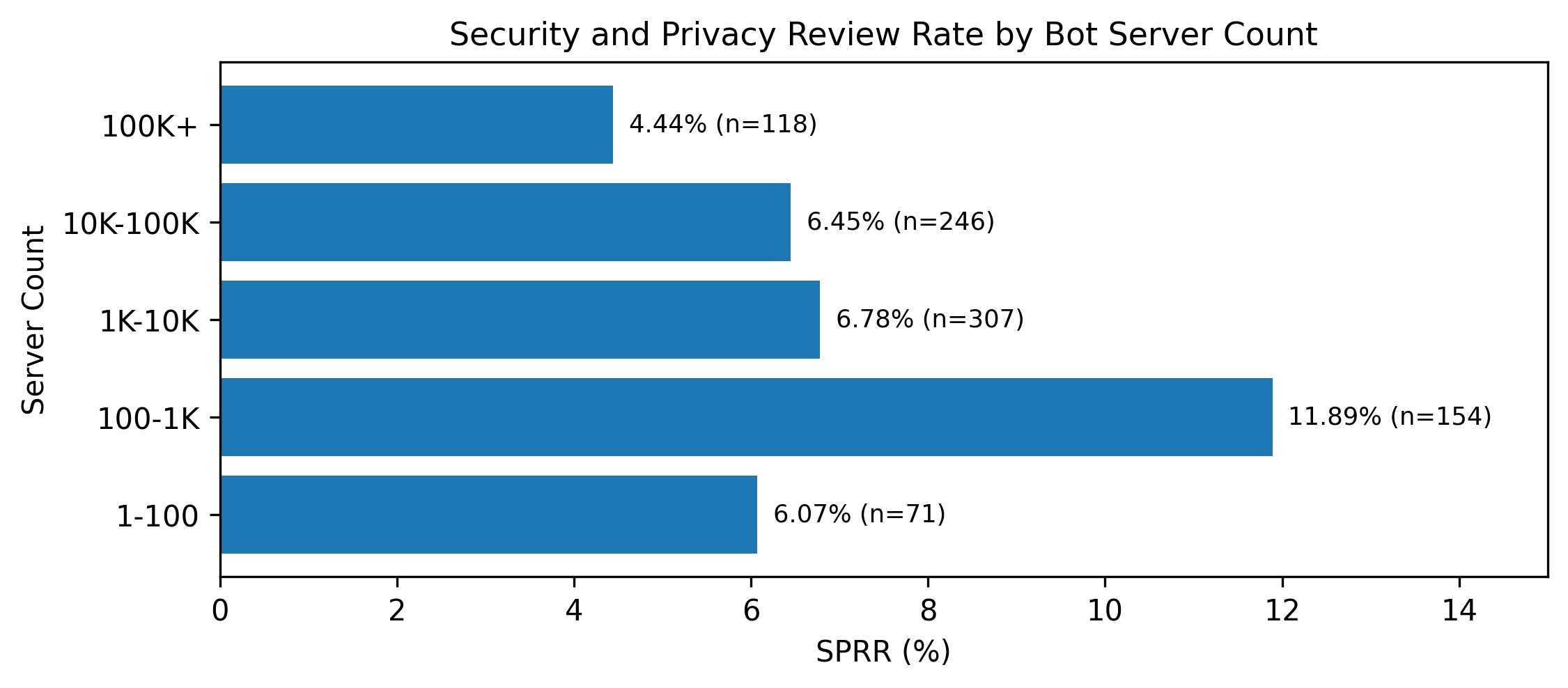}
    \caption{SPRR by bot server count. Mid-sized bots (100 to 1K servers) show the highest SPRR at 11.89\%.}
    \label{fig:rq4_server_count}
\end{figure}

Figure~\ref{fig:rq4_server_count} shows SPRR across five server-count bins for the 896 bots with a disclosed server count. Bots in the 100 to 1K range have the highest SPRR at 11.89\% (n=154), nearly double the rates of the other bins. Bots in the 1K to 10K range (6.78\%, n=307) and the 10K to 100K range (6.45\%, n=246) show moderate SPRR. The smallest bots (1--100 servers, 6.07\%, n=71) and the largest bots (100K+ servers, 4.44\%, n=118) have the lowest rates.

The spike in the 100 to 1K range may reflect a critical growth phase: these bots have enough users to generate S\&P feedback but may not yet have invested in the S\&P practices, documentation, or transparency that larger bots tend to adopt. The lower SPRR for the largest bots (100K+ servers) could indicate that widely adopted bots have been more thoroughly vetted by the community or have responded to earlier S\&P feedback. These interpretations are tentative because the analysis is observational and does not establish a causal relationship between server count and S\&P discussion.

\subsubsection{Code Transparency}
\label{sec:transparency}
Bots without a linked public GitHub repository have a slightly higher SPRR than those with one, 5.81\% (n=1,221) versus 4.91\% (n=236). Figure~\ref{fig:rq4_github} in Appendix~\ref{app:transparency} compares bots that link a public GitHub repository against those that do not. Because SPRR is valence-neutral, this difference does not establish that bots linking public GitHub repositories raise fewer S\&P problems. It may reflect greater perceived trust or more S\&P-conscious development practices, but our data do not distinguish these explanations.

\subsubsection{Metadata Correlation Matrix}

\begin{figure}[ht]
    \centering
    \includegraphics[width=\columnwidth]{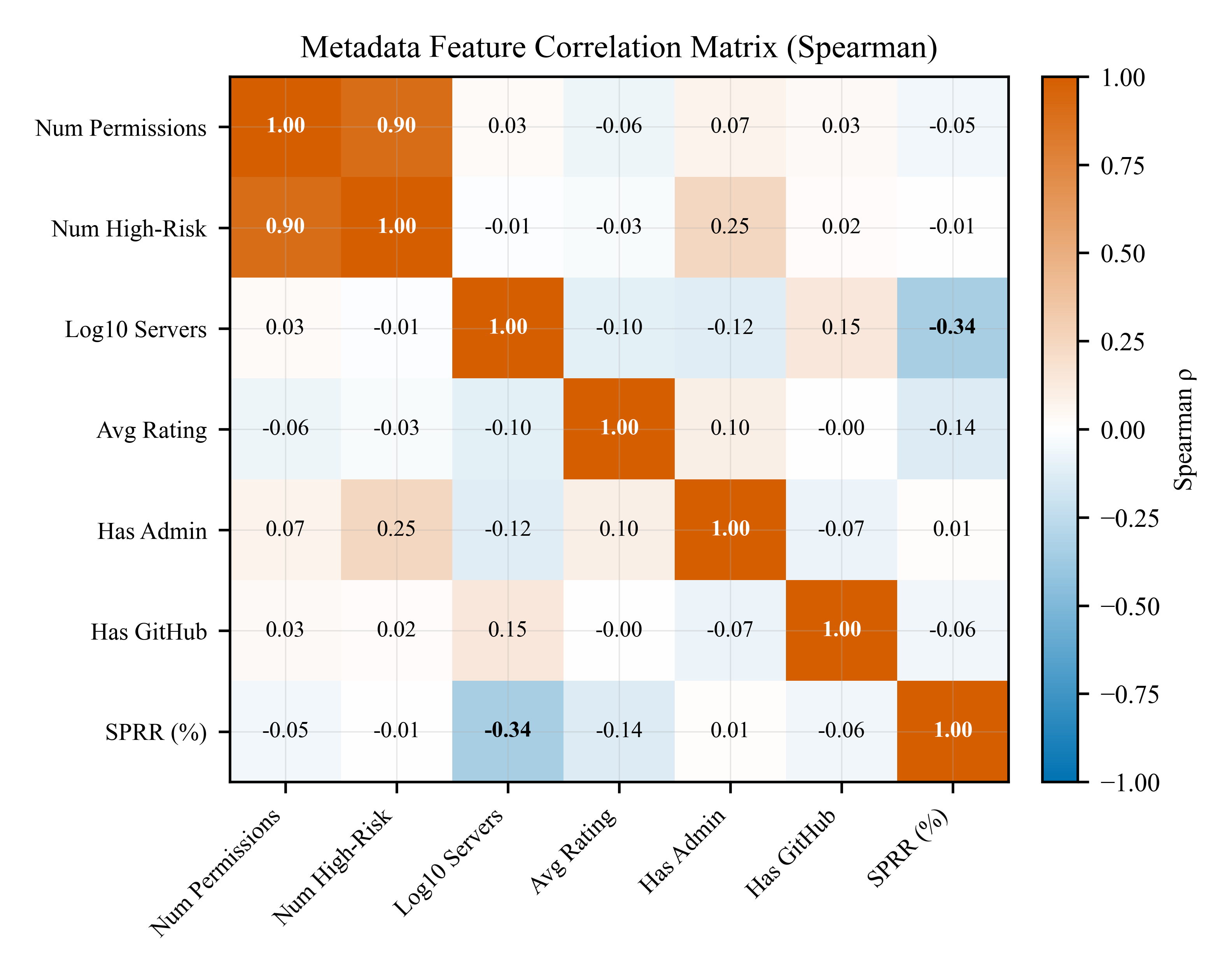}
    \caption{Spearman correlation matrix across bot metadata features and SPRR. Server count shows the strongest negative correlation with SPRR ($\rho=-0.34$).}
    \label{fig:rq4_correlation}
\end{figure}

Figure~\ref{fig:rq4_correlation} shows that server count has the strongest association with SPRR ($\rho=-0.34$). Other correlations are weak: number of permissions ($\rho=-0.05$), high-risk permissions ($\rho=-0.01$), average rating ($\rho=-0.14$), admin status ($\rho=0.01$), and GitHub availability ($\rho=-0.06$). Permission count and high-risk permission count are strongly correlated ($\rho=0.90$), indicating redundancy. These results suggest that visible metadata are weak predictors of S\&P discussion, except for the moderate negative association with server count.

We also examined privacy disclosures in Top.gg listings. A case-insensitive search for ``privacy policy'' in bots' short and long descriptions found references in only 177 of 23,102 bots (0.77\%); 99.23\% contained no such reference, indicating limited disclosure of bot data practices.
\section{Auditing Framework for Community Bots}
\label{sec:botlens}
To show that S\&P signals from user reviews along with bot metadata can be translated into a practical tool, we built \textbf{BotLens}, a proof-of-concept (PoC) auditing framework for community bots on social platforms.

\subsection{Motivation}
\label{sec:rq4_motivation}
Many users do not read reviews~\cite{camilleri2021who}. Even for users who read reviews, there are still risks. A low SPRR across our dataset means that a user needs to scroll through many reviews before hitting one that reveals an S\&P concern, assuming uniform distribution. In practice, platforms sort reviews by recency or helpfulness instead of topic, making it easy to miss S\&P-relevant reviews. This motivates the core design principle behind BotLens: rather than expecting users to find and interpret S\&P signals on their own, BotLens performs automated classification, aggregation, and summarization of review-derived signals.

\subsection{Framework Design}
\label{sec:rq4_design}
\subsubsection{Offline classification pipeline}
The pipeline from Section~\ref{sec:methodology} processes all 183,142 reviews: scraping from Top.gg, text cleaning, labeling via Ollama Llama 3 8B, DeBERTa classification, taxonomy mapping, affective-expression classification with GoEmotions, permission decoding, and statistical aggregation. The output is a pre-classified dataset where every review carries a binary S\&P label and a set of taxonomy tags, and every bot has a precomputed SPRR, permission profile, and sentiment composition among its S\&P-relevant reviews.

\subsubsection{MCP query server}
BotLens exposes six MCP tools over the pre-classified dataset:
\begin{enumerate}
    \item \texttt{get\_bot\_signals}: SPRR, star rating, requested permissions, reported S\&P themes, and sentiment composition for a named bot.
    \item \texttt{get\_taxonomy\_breakdown}: frequency distribution of taxonomy themes in a bot's S\&P reviews.
    \item \texttt{get\_review\_samples}: representative S\&P-relevant review text.
    \item \texttt{compare\_bots}: side-by-side comparison of two or more bots.
    \item \texttt{list\_high\_salience\_bots}: top-$k$ bots ranked by SPRR.
    \item \texttt{scan\_server\_bots}: accepts a bot manifest and returns a server-wide comparative report.
\end{enumerate}

Because every classification decision is precomputed, tool calls reduce to indexed lookups. No model inference is performed at query time. A user asks a question in natural language and receives a structured summary of review-derived S\&P signals in milliseconds. The full algorithmic pipeline is detailed in Appendix~\ref{app:algorithm}.

\subsection{Use Cases in Practice}
\label{sec:rq4_usecases}
We present two representative BotLens use cases.

\subsubsection{Use Case 1 - Checking a specific bot}
\begin{figure}[h]
    \centering
    \includegraphics[
        width=\columnwidth,
        height=0.30\textheight,
        keepaspectratio
    ]{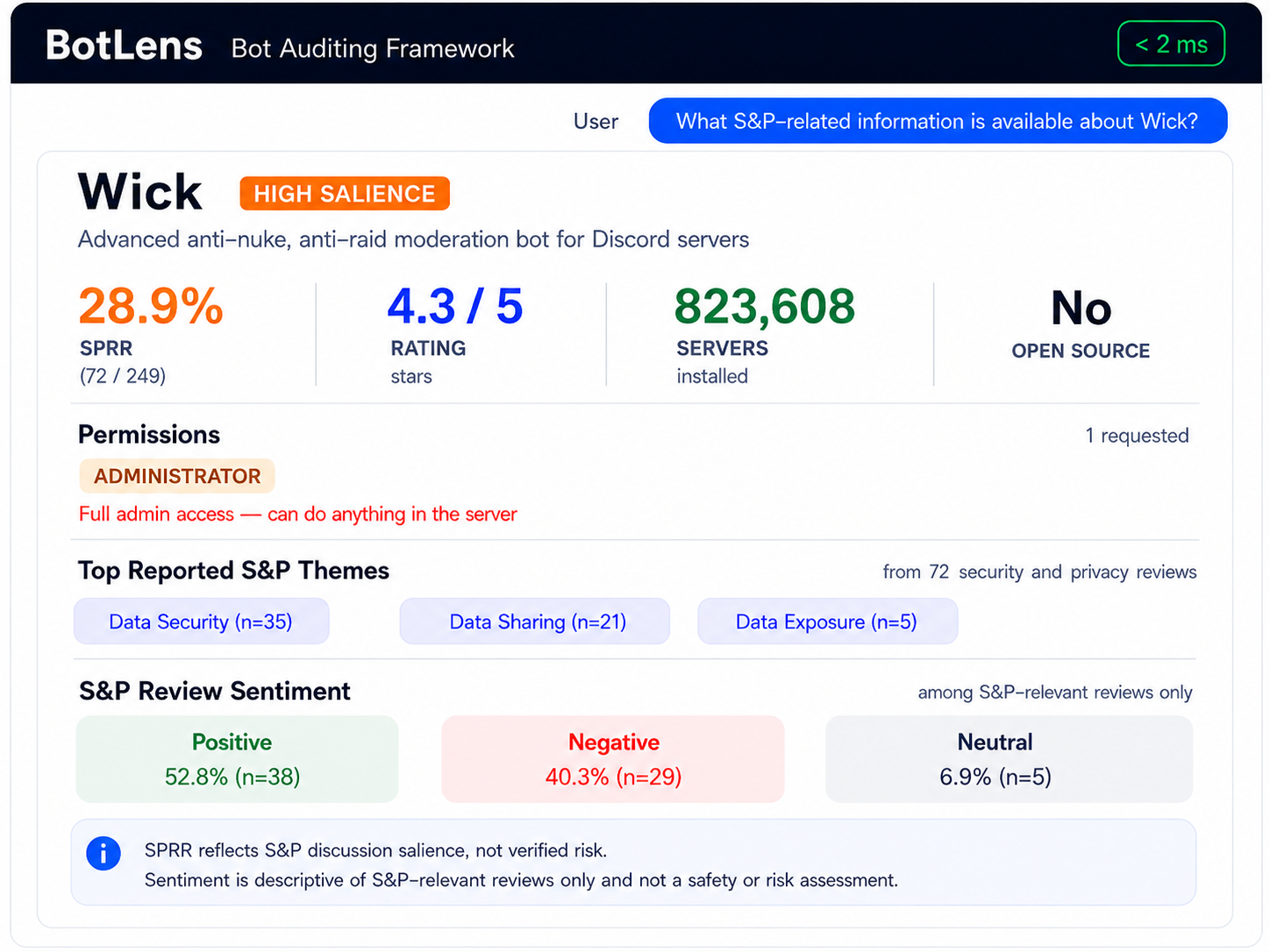}
    \caption{Use Case 1: Bot-level checkup. BotLens presents Wick's SPRR, requested permissions, reported S\&P themes, and sentiment composition.}
    \label{fig:rq5_uc1}
\end{figure}

Figure~\ref{fig:rq5_uc1} depicts an example of checking a bot named ``Wick.'' BotLens returns its SPRR, average rating, requested permissions, reported S\&P themes, and sentiment composition without requiring a platform login and in under 2\,ms. As shown, Wick has a 4.3-star rating and 28.9\% SPRR, requests \texttt{ADMINISTRATOR}-level access, and is most frequently associated with Data Security, Data Sharing, and Data Exposure. Among its 72 S\&P reviews, 52.8\% are positive, 40.3\% negative, and 6.9\% neutral. These signals provide S\&P information not captured by the aggregate rating alone.

\subsubsection{Use Case 2 - Auditing an entire server}
\begin{figure}[h]
    \centering
    \includegraphics[
        width=\columnwidth,
        height=0.30\textheight,
        keepaspectratio
    ]{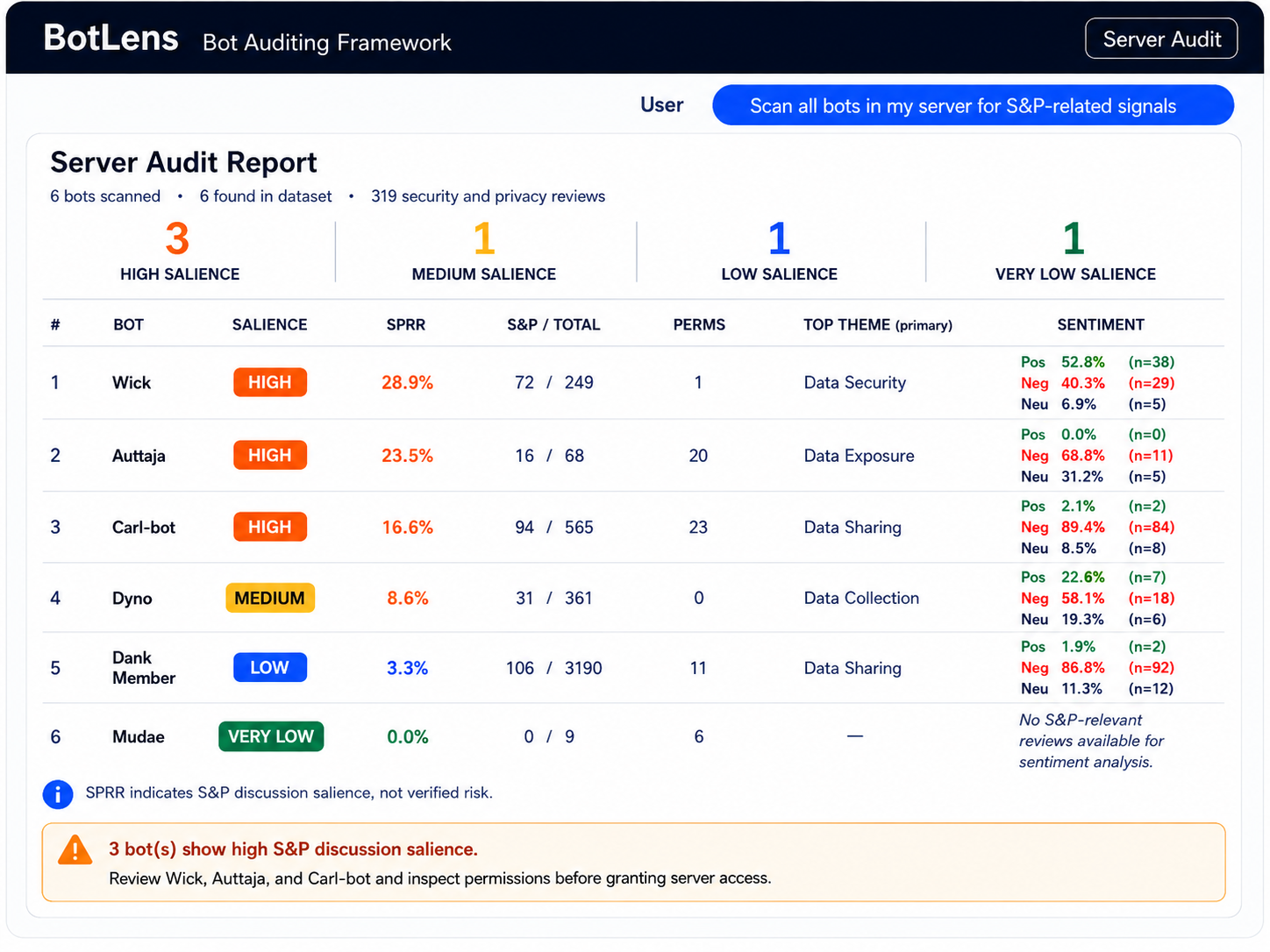}
    \caption{Use Case 2: Server-wide audit comparing SPRR, reported S\&P themes, requested permissions, and sentiment composition.}
    \label{fig:rq5_uc2}
\end{figure}

BotLens provides a server administrator a way to quickly check all the bots installed on a server. Particularly, BotLens searches for \texttt{APP}-badged members, paginates through the results, and exports a bot manifest. The manifest is then processed by \texttt{scan\_server\_bots}, which returns a comparative report containing SPRR, reported S\&P themes, requested permissions, and sentiment composition. Figure~\ref{fig:rq5_uc2} shows an example of comparing six bots in a server. Three bots have an SPRR above 10\%, indicating that S\&P topics appear comparatively frequently in their reviews. The report presents sentiment composition alongside SPRR so that an administrator can distinguish bots whose S\&P discussion is predominantly critical from those where it is predominantly positive.

\subsection{Time Performance}
\label{sec:rq4_performance}
Across 22 queries, \texttt{get\_bot\_signals} (11 calls), \texttt{get\_taxonomy\_breakdown} (1 call), and \texttt{get\_review\_samples} (4 calls) averaged 1.7\,ms, 1.6\,ms, and 1.9\,ms, respectively. \texttt{scan\_server\_bots} (6 calls) averaged 14.7\,ms and peaked at 31.3\,ms, with a session-wide mean of 5.0\,ms.

\section{Discussion}
\label{sec:discussion}

In this section, we discuss the implications for platform designers, bot developers, and end users, then address limitations and future directions.

\subsection{Key Findings and Implications}

\subsubsection{S\&P discussion is meaningful but unevenly distributed}

The observed pipeline yield of 4.82\% indicates that nearly 1 in 20 bot reviews was identified and retained as S\&P-relevant.
This is a substantive review-derived signal, especially because Top.gg review writers are not specifically asked about S\&P. The signal is not uniform across the ecosystem.
Confession and anonymity bots exceed 50\% SPRR, meaning that our pipeline identified S\&P topics in more than half of their reviews.
This variation suggests that platform disclosures should account for bot function and the sensitivity of the data handled.

\subsubsection{Bot function shapes the disposition of S\&P concerns}

SPRR variation across bot categories suggests that S\&P discussion is associated with bot functionality and data sensitivity, though these relationships are associative rather than causal. Moderation and governance bots may raise concerns about surveillance, permissions, exclusion, and misuse because they monitor activity and affect participation. Such bots should follow least privilege, disclose monitored activities, explain automated decisions, and provide appeal mechanisms. Confession and anonymity bots raise different concerns because they process sensitive disclosures and identity information. They should disclose whether submitter identities can be recovered, how submissions and metadata are stored, and when they are deleted. Platforms should also provide function-specific notices to affected server members, rather than only to installing administrators. The weak association between permission metadata and SPRR may reflect a privacy-paradox pattern in which users tolerate S\&P concerns because of functionality, community dependence, or limited alternatives. Because our review data do not observe installation or continued-use behavior, this remains a hypothesis for future user studies.

\subsubsection{Star ratings do not capture S\&P discussion salience}
Star ratings summarize multiple dimensions of user experience, including functionality, reliability, and ease of use, and therefore may obscure S\&P-related feedback. For example, bots such as Wick (SPRR = 28.9\%, stars = 4.3) and Auttaja (SPRR = 23.5\%, stars = 3.3) have competitive average ratings while also receiving substantial S\&P-related discussion. These findings indicate that aggregate ratings should not be interpreted as endorsements of a bot's S\&P practices. Platforms could complement general star ratings with a separate indicator that summarizes the pipeline-identified rate and composition of S\&P-related reviews.

\subsubsection{Permission metadata is a weak standalone predictor}
RQ3 shows weak associations between user-reported S\&P concerns and permission count, administrator status, or individual high-risk permissions. Admin and non-admin bots have nearly identical SPRR (5.69\% vs.\ 5.64\%), suggesting that broader permission scope does not necessarily correspond to more S\&P discussion. RQ1 instead indicates that bot function and data sensitivity matter more. For example, confession bots may request few permissions while processing highly sensitive disclosures. Permission auditing alone therefore provides an incomplete picture.

\subsubsection{User reviews capture what policies miss}

Prior work found that only 4.35\% of Discord chatbots that request permissions provide a privacy policy~\cite{edu2022exploring}; our dataset shows an even lower rate: just 0.77\% of bots reference a privacy policy in their Top.gg listings. Reviews can report perceived or experienced bot behaviors that may not appear in formal policies. Review mining can therefore complement policy analysis on platforms that lack consistent disclosure requirements, although review claims remain self-reported and are not independently verified.

\subsubsection{Affective-expression patterns complement star ratings}

RQ2 shows that affective-expression patterns broadly align with the polarized star-rating distribution. Neutral and Joy are common across several S\&P themes, whereas Anger and Confusion appear more frequently in reviews concerning Data Exposure and DM Privacy. These findings describe affective language in review text and should not be interpreted as direct evidence of users' underlying emotions or as proof of binary S\&P judgments.

\subsection{Recommendations}

\subsubsection{For Social Platforms with Community-based Bots}

Platforms should consider presenting a dedicated S\&P review indicator alongside star ratings, analogous to mobile-app privacy disclosures~\cite{apple2020privacydetailslabel,google2022datasafety}. Such an indicator could report SPRR as the pipeline-identified rate of S\&P-related discussion and distinguish reviews that express concerns from those that provide positive feedback. Platforms could also provide category-specific disclosures for bots that handle sensitive content, such as confession, anonymity, and anti-nuke bots, without treating elevated SPRR as verified evidence of risk.

\subsubsection{For Bot Developers} 
Developers should request only necessary permissions and clearly explain their bots' data access. The Positive Privacy theme (2.2\% of S\&P reviews) suggests that users reward transparency and privacy-protective features.

\subsubsection{For Administrators and Users}
Server administrators should examine both requested permissions and review-derived S\&P signals when evaluating bots. BotLens supports server-wide comparison by surfacing SPRR, reported S\&P themes, sentiment composition, and requested permissions for installed bots. These outputs can help identify bots that warrant closer examination. The comparatively high SPRR among bots deployed in 100--1K servers also motivates closer examination of this range.

\subsection{Limitations and Future Work}
\subsubsection{Data Source}
Our data are limited to bots on Top.gg, and user reviews are self-reported. Users who write reviews may not represent the broader population, and privacy-concerned users may be overrepresented, which could increase the observed pipeline yield. We did not verify whether the claims in individual reviews are factually correct. We could not verify claims retrospectively because historical bot versions, server configurations, logs, and incident context were unavailable. Our findings therefore characterize review writers' reported experiences and perceptions rather than confirmed bot behavior or verified S\&P violations. Future work can examine their factual basis through longitudinal, dynamic analysis.

\subsubsection{Manual Validation and SPRR Interpretation}
\label{sec:limitations}

Both the teacher model validation (Section~\ref{sec:irr}) and S\&P candidate validation (Section~\ref{sec:stage3a}) used the same S\&P relevance criteria through our 31-theme taxonomy. The teacher model validation involved open-ended labeling of a stratified sample without access to the teacher labels. The candidate validation focused on classifier-selected reviews enriched for predicted S\&P relevance. The focused context here may have made subtle S\&P content easier to recognize. On the other hand, the open-ended labeling in the teacher model validation was not sensitive to such subtle cases, although the annotators agreed strongly. This provides a plausible explanation for the positive-rate difference between the two validation procedures. Cohen's \(\kappa\) and raw agreement establish labeling consistency within the sample, but not labeling sensitivity, dataset completeness, corpus prevalence, or ground-truth status. The 0.30\% teacher-positive rate describes the teacher labels within the 54,942-review subset and is not a corpus-prevalence estimate. The 4.82\% SPRR represents the observed pipeline yield, not the true prevalence of S\&P relevance.

\subsubsection{Generalizability Beyond Discord}

The methodological components of this work are platform-independent. The SPRR/TPR metrics and the BotLens auditing workflow are applicable to other bot marketplaces, messaging-platform app directories, and AI-agent ecosystems. Several findings rest on conditions that are not specific to Discord. In particular, star ratings summarize many dimensions of user experience on any review platform, so their tendency to obscure S\&P-related feedback is not Discord-specific. The same applies to the weak association between requested permissions and where pipeline-identified S\&P discussion concentrates. Permission models describe capability rather than the sensitivity of the content a bot handles, which is common across app stores and bot platforms.

Some results are specific to our setting. The pipeline-identified SPRR values, including the 4.82\% ecosystem-wide pipeline yield and the rates above 50\% for confession and anonymity bots, reflect Top.gg's reviewer population and bot composition and would likely differ elsewhere. The nine Discord-specific themes are tied to platform mechanisms such as Gateway Intents, webhooks, and administrator-mediated installation. Platforms without these mechanisms would require different extensions, although the four theme groups and the 20 themes inherited from prior work~\cite{akgul2024decade} remain applicable. Discord's combination of community scale with server-wide installation by an administrator also differs from platforms where bots operate in private workspaces~\cite{slack2024apps}.

\subsubsection{LLM for Labeling and Taxonomy Assignment}
Residual labeling errors may propagate into downstream analyses. The classifier evaluation reported in Section~\ref{sec:stage2c}
rests on a small number of positive examples, and the human annotations in Section~\ref{sec:irr} show substantial but imperfect agreement. Language identification and translation were automated. Short,
informal, or code-switched reviews may be assigned an incorrect language, and translation errors may propagate into S\&P classification, taxonomy assignment, and affective-expression analysis. The taxonomy assignments and affective-expression labels were not independently evaluated against a dedicated ground truth. We therefore interpret the resulting distributions as model-assisted descriptive findings and acknowledge that errors may remain. Future work can explore stronger models and multi-pass/multi-label taxonomy assignment.

\subsubsection{Real-Time and In-Situ Auditing}
BotLens is a PoC to demonstrate feasibility, not completeness. Its outputs summarize SPRR, sentiment composition, reported S\&P themes, and requested permissions, but they do not verify the claims in the reviews or determine if a bot is safe. We did not evaluate how BotLens improves decision-making or how users respond to S\&P signals at installation. Future work can evaluate BotLens during bot-installation decisions.

\section{Conclusion}

In this paper, we present a large-scale study of user-reported S\&P concerns in the Discord bot ecosystem, which offers a baseline for S\&P auditing of community-based bots on social platforms. We contribute a 31-theme S\&P taxonomy and BotLens, an auditing framework that translates review-derived signals into on-demand decision support. Our findings show that S\&P discussion is especially prevalent among confession and anonymity bots. We also find that metadata features such as permission count and administrator status have only weak associations with user-reported S\&P concerns. These results indicate that bot function and the sensitivity of the content a bot handles are more informative than permission scope alone.

\section*{Ethical Considerations}
\label{sec:ethics}

This section discusses the ethics of our work.

\subsection{Stakeholder Identification and Impact Analysis}

The stakeholders of this work include (1) the research team, (2) Discord bot users, (3) Discord server administrators, (4) third-party bot developers, (5) Discord and bot-listing platforms, and (6) the security and privacy research community. We discuss the potential impact on each group below.

\textbf{\textit{Research Team:}}
We did not recruit participants, interact with users, or collect data from private Discord servers. We did not access direct messages, server logs, or user accounts. Our analyses used publicly available bot metadata and user reviews from Top.gg. The analyzed materials consisted of publicly visible review text, bot descriptions, permissions, tags, ratings, and other listing-level information. Because public review text may still contain usernames, personal references, or other potentially identifiable information, we excluded such details from the paper and reported findings primarily in aggregate. Our institution's Institutional Review Board (IRB) formally determined that IRB review was not required for this study under DHHS and FDA regulations.

\textbf{\textit{Discord Bot Users:}}
Bot users are the people this work is most directly intended to help. When a bot is added to a server, the average member has little control over what it can access or do. Our study analyzes reviews that users have already posted publicly to surface the security and privacy concerns they raise without researcher prompting. The aim is to make those concerns easier to identify and act upon. We report findings in aggregate and avoid drawing attention to individual review writers.

\textbf{\textit{Discord Server Administrators:}}
Server administrators install bots, and that decision can affect everyone else in the server. This work supports administrators by providing clearer review-derived S\&P signals when star ratings or deployment counts are uninformative about S\&P practices. BotLens can summarize each installed bot's SPRR, reported S\&P themes, sentiment composition, and requested permissions, helping administrators identify bots that warrant closer examination without treating those signals as verified evidence of risk or misconduct.

\textbf{\textit{Third-Party Bot Developers:}}
Developers may face reputational harm because individual bots are named in the paper. A reported pattern could be interpreted as an accusation even though our analysis does not establish misconduct. To reduce this risk, we distinguish user-reported concerns and review patterns from verified bot behavior. Where we name a bot, we do so to illustrate a pattern, not to single out a developer. We frame the results as signals that may warrant further examination, not as verdicts.

\textbf{\textit{Discord and Bot Marketplaces:}}
Our findings identify gaps in how bot platforms communicate security and privacy information to users. Platforms could use these findings to improve permission disclosures, surface privacy policies more clearly, organize reviews more effectively, or provide stronger decision support. We did not bypass platform protections or collect information from authenticated or private pages. Our crawler accessed public pages and used rate limits to reduce load on Top.gg.

\textbf{\textit{Security and Privacy Research Community:}}
For the research community, this work contributes a method, taxonomy, and dataset for studying user-reported security and privacy concerns in community bot ecosystems at scale. The approach may also be applicable to other bot marketplaces, messaging platforms, or AI-agent ecosystems where users provide public feedback.

\subsection{Potential Harm and Mitigation}

First, public reviews may still contain sensitive or identifiable information. We do not publish personally identifiable information from reviews, and we report findings in aggregate.

Second, the work may create reputational risk for bot developers. A bot with a high Security and Privacy Review Rate (SPRR) is not necessarily unsafe or malicious. SPRR measures the proportion of reviews identified and retained as S\&P-relevant by our pipeline. It does not measure the factual correctness of the reported claims or the underlying technical behavior of a bot. We therefore frame the findings as observations about user-reported experiences and perceptions rather than as accusations.

Third, there is a risk of overgeneralization. A high SPRR means that our pipeline identifies S\&P topics more frequently in reviews of that bot or category, while a low SPRR does not guarantee a bot is safe. We treat BotLens as a decision-support tool, not a verdict. Automated classification may also introduce errors. We report the classifier's measured performance and the uncertainty around it (Section~\ref{sec:stage2c}), manually validated the full set of classifier-positive reviews (Section~\ref{sec:stage3a}), and acknowledge that residual error may remain.

\subsection{Decision and Research Justification}

The motivation for this work is the gap between what bots can do inside a server and what the people in that server can observe. A bot installed by one administrator can affect other members who may not know what permissions were granted or how their data may be handled. Our study shows that public reviews contain security and privacy concerns that are often not captured by the signals users commonly rely on, such as star ratings, deployment counts, or permission lists.

We weighed the contribution of this research against the potential costs to users, platforms, and developers, and we believe the balance supports conducting this work. Users, server administrators, developers, and platforms may benefit from a clearer understanding of where pipeline-identified security and privacy discussion is concentrated, while BotLens demonstrates how these review-derived signals can be made practically accessible. We minimized potential harm by using only public data, excluding identifying details, aggregating results, distinguishing reported concerns from verified behavior, and presenting the results with appropriate caution.

\section*{Acknowledgment}
We would like to thank the anonymous reviewers for their time and constructive feedback. We also thank Long Huynh for his help in the initial phase of the project.



\bibliographystyle{IEEEtran}
\bibliography{IEEEabrv,references}
%




\appendix
\raggedbottom

\subsection{BotLens Auditing Algorithm}
\label{app:algorithm}

Algorithm~\ref{alg:botlens} details the end-to-end auditing procedure summarized in Section~\ref{sec:botlens}. Phase~2 computes the SPRR values defined in Section~\ref{sec:analysis}.

\begin{algorithm}[!hbp]
\caption{BotLens End-to-End Bot Auditing Algorithm}
\label{alg:botlens}
\begin{algorithmic}[1]
\REQUIRE Target server URL $u$, classified dataset $\mathcal{D}$, taxonomy $\mathcal{T}$
\ENSURE Ranked S\&P signal report $\mathcal{R}$
\medskip
\STATE \textbf{// Phase 1: Bot Discovery}
\STATE Launch headless browser in the administrator's own authenticated session
\STATE Navigate to $u$; query for \texttt{authorType: bot}
\STATE $B \leftarrow \emptyset$
\REPEAT
    \FOR{\textbf{each} member $b$ on current results page}
        \IF{$b$ has the \texttt{APP} badge}
            \STATE $B \leftarrow B \cup \{b.\textit{username}\}$
        \ENDIF
    \ENDFOR
    \STATE Advance to next results page
\UNTIL{no further pages remain}
\STATE Export $B$ to bot manifest file

\medskip
\STATE \textbf{// Phase 2: S\&P Signal Aggregation}
\STATE $\mathcal{R} \leftarrow \emptyset$

\FOR{\textbf{each} bot $b \in B$}
    \STATE $\mathcal{D}_b \leftarrow$ reviews for $b$ in $\mathcal{D}$

    \IF{$\mathcal{D}_b \neq \emptyset$}
        \STATE $P \leftarrow \{r \in \mathcal{D}_b \mid r.\textit{is\_sp} = 1\}$
        \STATE $\textit{SPRR}_b \leftarrow (|P| \;/\; |\mathcal{D}_b|) \times 100$
        \STATE $s_b \leftarrow
        \langle
        \textit{SPRR}_b,\;
        \bar{\textit{rating}}_b,\;
        \textit{sentiment}_b,\;
        \textit{perms}_b
        \rangle$
    \ELSE
        \STATE $s_b \leftarrow \textit{PermissionAudit}(b)$
        \COMMENT{metadata fallback}
    \ENDIF

    \STATE $s_b.\textit{themes} \leftarrow
    \textit{TaxonomyFlags}(b,\mathcal{T},\mathcal{D}_b)$

    \STATE $\mathcal{R} \leftarrow \mathcal{R} \cup \{s_b\}$
\ENDFOR

\RETURN $\mathcal{R}$ sorted by $\textit{SPRR}$ discussion salience descending

\medskip
\STATE \textbf{// Subroutine: TaxonomyFlags($b,\mathcal{T},\mathcal{D}_b$)}
\STATE $F \leftarrow \emptyset$

\FOR{\textbf{each} theme $c \in \mathcal{T}$}
    \STATE $n_c \leftarrow
    |\{r \in \mathcal{D}_b
    \mid r.\textit{is\_sp}=1
    \land c \in \textit{Themes}(r)\}|$

    \IF{$n_c > 0$}
        \STATE $F \leftarrow F \cup \{(c,n_c)\}$
    \ENDIF
\ENDFOR

\RETURN $F$ sorted by frequency descending
\end{algorithmic}
\end{algorithm}

\subsection{Affective Expressions by Language}
\label{app:language_emotions}

\begin{figure}[H]
    \centering
    \includegraphics[width=0.9\columnwidth]{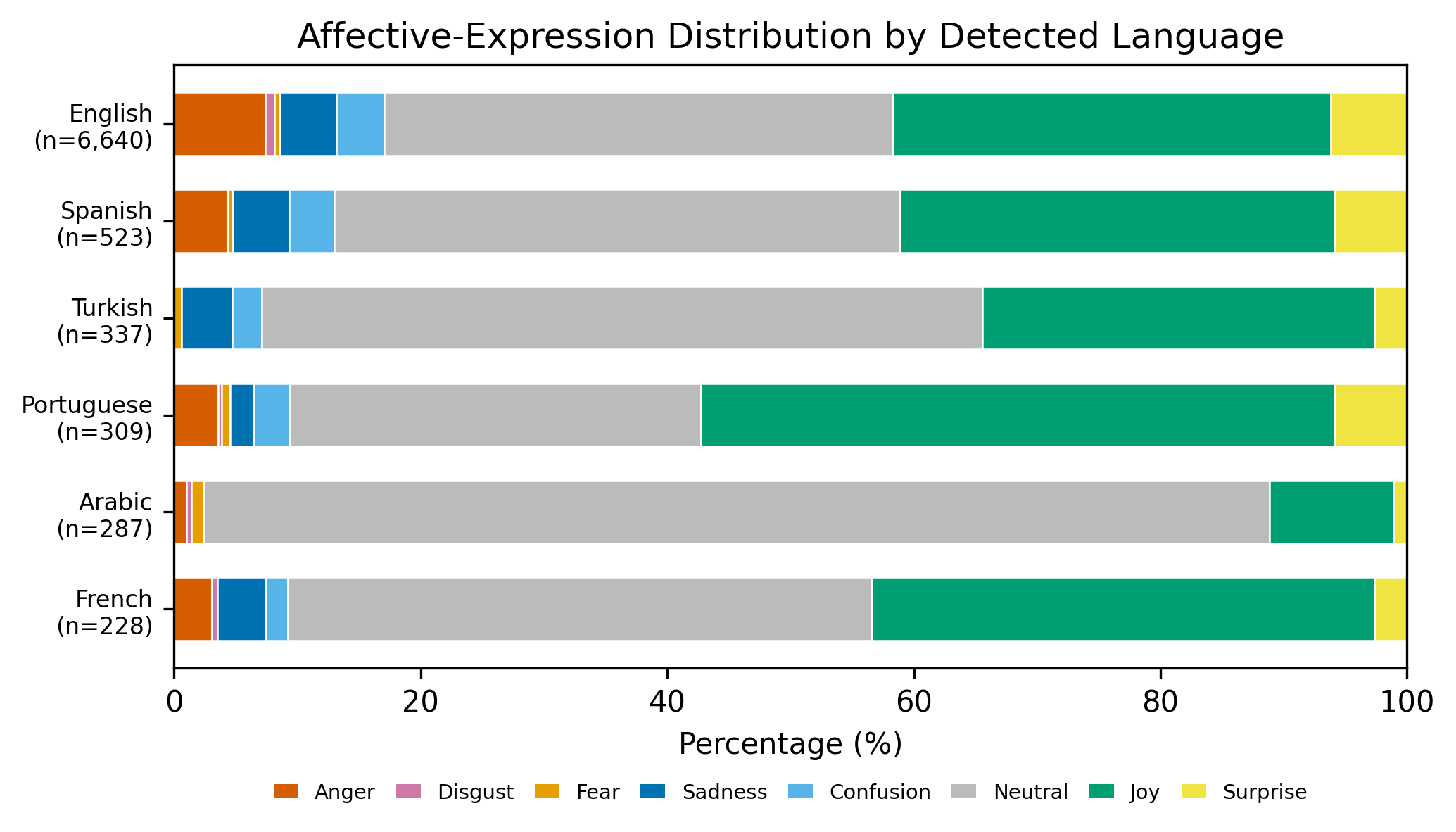}
    \caption{Affective-expression distribution in S\&P-relevant reviews by detected language. Languages with at least 100 S\&P-relevant reviews are shown.}
    \label{fig:language_emotions}
\end{figure}

Figure~\ref{fig:language_emotions} shows the distribution of affective-expression labels across the six detected languages with at least 100 S\&P-relevant reviews (see Table~\ref{app:languages} for the full language distribution). Together, these languages account for 94.2\% of the S\&P-relevant set. Neutral and Joy are the two largest categories in all six language groups, consistent with the theme-level and category-level patterns reported in Section~\ref{sec:rq2}. English- and Spanish-language reviews show similar profiles, with Neutral accounting for 41\% and 46\%, respectively, and Joy accounting for 35\% in both. French-language reviews are comparable at 47\% Neutral and 41\% Joy. Turkish-language reviews contain a larger Neutral share at 58\%, Portuguese-language reviews show the highest Joy share at 51\% alongside the lowest Neutral share at 33\%, and Arabic-language reviews show the highest Neutral share at 86\%. The relative shares therefore vary across languages, but the ordering does not: matter-of-fact and positive labels predominate over negative ones in every group. Because GoEmotions was applied to machine-translated text for all non-English reviews, we treat the differences in magnitude as model-assigned and do not interpret them as differences in how review writers express S\&P concerns.

\begin{table}[H]
\centering
\caption{Language distribution of Discord bot reviews. The table reports the top 25 of 49 valid detected languages, the remaining 24 valid languages, and reviews with undetermined or invalid language labels.}
\label{app:languages}
\scriptsize
\begin{tabular}{rlrr}
\toprule
\textbf{\#} & \textbf{Language} & \textbf{Count} & \textbf{\%} \\
\midrule
1  & English        & 100,711 & 49.92 \\
2  & Turkish        &  10,472 &  5.19 \\
3  & Spanish        &   9,342 &  4.63 \\
4  & Portuguese     &   7,382 &  3.66 \\
5  & Somali         &   6,046 &  3.00 \\
6  & Afrikaans      &   4,657 &  2.31 \\
7  & French         &   3,526 &  1.75 \\
8  & German         &   3,119 &  1.55 \\
9  & Arabic         &   2,691 &  1.33 \\
10 & Romanian       &   2,450 &  1.21 \\
11 & Dutch          &   2,374 &  1.18 \\
12 & Italian        &   2,346 &  1.16 \\
13 & Welsh          &   2,201 &  1.09 \\
14 & Filipino       &   2,027 &  1.00 \\
15 & Slovenian      &   1,958 &  0.97 \\
16 & Catalan        &   1,906 &  0.94 \\
17 & Polish         &   1,869 &  0.93 \\
18 & Norwegian      &   1,805 &  0.89 \\
19 & Swahili        &   1,649 &  0.82 \\
20 & Vietnamese     &   1,626 &  0.81 \\
21 & Indonesian     &   1,618 &  0.80 \\
22 & Danish         &   1,456 &  0.72 \\
23 & Croatian       &   1,329 &  0.66 \\
24 & Czech          &   1,203 &  0.60 \\
25 & Estonian       &   1,069 &  0.53 \\
\midrule
   & \textit{Other (24 valid languages)} & 6,665 & 3.30 \\
   & \textit{Undetermined/invalid} & 18,236 & 9.04 \\
   & \textbf{Total} & \textbf{201,733} & \textbf{100.00} \\
\bottomrule
\end{tabular}
\end{table}

\subsection{SPRR by Code Transparency}
\label{app:transparency}

\begin{figure}[!hb]
    \centering
\includegraphics[width=0.85\columnwidth]{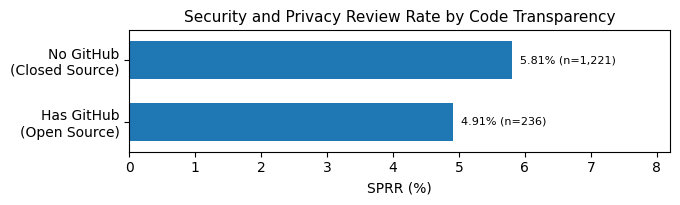}
    \caption{SPRR by code transparency. Closed-source bots (5.81\%) have a slightly higher security and privacy review rate than open-source bots (4.91\%).}
    \label{fig:rq4_github}
\end{figure}

Figure~\ref{fig:rq4_github} compares SPRR for bots that link a public GitHub repository against those that do not, supporting the code-transparency result reported in Section~\ref{sec:transparency}.

\subsection{Complete Security and Privacy Taxonomy}
\label{app:taxonomy}

Table~\ref{tab:taxonomy} presents the full 31-theme taxonomy developed in Section~\ref{sec:taxonomy}. The 11 themes marked \textsc{New} were derived through the inductive coding described in that section.

\begin{table*}[p]
\centering
\caption{Complete Security and Privacy Taxonomy: 31 themes with ${\sim}$139 fine-grained concepts. Themes marked \textsc{Original} are preserved from prior work~\cite{akgul2024decade}; those marked \textsc{New} were developed in our study.}
\label{tab:taxonomy}
\footnotesize
\renewcommand{\arraystretch}{1.1}
\begin{tabular}{p{0.14\textwidth} p{0.06\textwidth} p{0.72\textwidth}}
\toprule
\textbf{Theme} & \textbf{Source} & \textbf{Fine-Grained Aspects} \\
\midrule

\multicolumn{3}{l}{\cellcolor[gray]{0.9}\textbf{Theme Group 1: Data Practices (7 themes)}} \\
\midrule

Data Collection & \textsc{Original} & Collecting unnecessary personal data; Purpose of data collection; Data aggregation; Data minimization \\

Data Sharing & \textsc{Original} & Accidental sharing; Forced sharing; Unintended sharing; Cross-app sharing; Secondary use; Calendar sharing; Password protected sharing \\

Data Deletion & \textsc{Original} & Browsing history removal; Delete video history; Delete search history; Cookie removal; Photo deletion; Video deletion; File deletion; Profile deletion \\

Remove PII & \textsc{Original} & Remove name; Remove address; Remove phone number; Remove personal information; Remove photo \\

Data Exposure & \textsc{Original} & To advertisers; To app developers; Public accessibility; Data disclosure; Safety; Use limitation \\

Data Hiding & \textsc{Original} & Name; Video; Events; Albums; Photos; Contacts; Display name; Email; Messages; Notes; Playlist; Folders; Meetings \\

Location \& Tracking & \textsc{Original} & Location data; Tracking \\

\midrule
\multicolumn{3}{l}{\cellcolor[gray]{0.9}\textbf{Theme Group 2: User Rights \& Controls (7 themes)}} \\
\midrule

Consent & \textsc{Original} & Giving consent; Agreement; Authorize; Consent process; Opt-out (notice awareness); Without consent; Mis-activation; Forced consent \\

Privacy Controls & \textsc{Original} & Privacy settings; Changing personal information; Location sharing; Download my data; Parental controls; Privacy defaults; Password protected controls \\

Anonymity / Identification & \textsc{Original} & User wants to be anonymous; Incognito; Hide group participants; Fear of identification; Misattribution \\

Advertising & \textsc{Original} & Ads personalization; Paid services \\

Privacy Policies \& Laws & \textsc{Original} & \emph{(General concept; no sub-aspects in original)} \\

Privacy Invasion & \textsc{Original} & \emph{(General concept; no sub-aspects in original)} \\

Positive Privacy & \textsc{Original} & \emph{(General concept; no sub-aspects in original)} \\

\midrule
\multicolumn{3}{l}{\cellcolor[gray]{0.9}\textbf{Theme Group 3: Security \& Safety (6 original + 2 new themes)}} \\
\midrule

Data Security & \textsc{Original} & Safety; Protection; Breach of confidentiality; Data breach; Account hacking; Fake profiles \\

Password Issues & \textsc{Original} & \emph{(General concept; no sub-aspects in original)} \\

Data Accuracy & \textsc{Original} & Inaccurate; Obsolete; Wrong association \\

Safety & \textsc{Original} & Blackmail; Appropriation; Intrusion; Decisional interference \\

Selling Data & \textsc{Original} & To 3rd parties; For ads purposes \\

Surveillance & \textsc{Original} & Spying; Stalkerware; Satellite \\

Phishing \& Scams & \textsc{New} & Fake Nitro gift links; Fake verification links; Token grabbing attempts; Credential harvesting; Scam links in bot messages \\

Data Retention & \textsc{New} & Message logs kept indefinitely; User data not deleted after leaving server; Data persists after bot is kicked; No data retention policy; Chat history stored permanently \\

\midrule
\multicolumn{3}{l}{\cellcolor[gray]{0.9}\textbf{Theme Group 4: Discord-Specific Risks (9 themes)}} \\
\midrule

Over-Listening & \textsc{New} & Reading messages in unrelated channels; Processing messages without being called; Background monitoring of all messages; Listening beyond stated scope; Monitoring activity when not invoked \\

Permission Scope Mismatch & \textsc{New} & Requesting admin permissions unnecessarily; Over-broad permission scopes; Permissions don't match bot functionality; Requesting dangerous permissions for utility bot; Music bot requesting message read access \\

Message Content Access & \textsc{New} & Bot reads all messages via \texttt{MESSAGE\_CONTENT} intent; Message content intent used unnecessarily; Bot can see private conversations in channels; Logging message content without disclosure \\

DM Privacy & \textsc{New} & Unsolicited DMs from bot; Bot logging DM conversations; Spam DMs after adding bot; Bot reads DMs without need; DM advertising/promotion \\

Gateway Intent Abuse & \textsc{New} & Tracking online/offline status via \texttt{GUILD\_PRESENCE}; Accessing full member list via \texttt{GUILD\_MEMBERS}; Using privileged intents beyond stated purpose; Monitoring user presence patterns \\

Webhook Misuse & \textsc{New} & Creating webhooks without permission; Impersonating users via webhooks; Bypassing server audit logs; Webhook spam \\

Server Profiling & \textsc{New} & Harvesting server member lists; Mapping server structure and roles; Collecting server analytics without consent; Exporting server data \\

Voice Privacy & \textsc{New} & Voice recording without consent; Voice data stored without notice; Music bot recording conversations; Voice activity monitoring; Voice data shared with third parties \\

Bot Moderation Transparency & \textsc{New} & No reason given for ban/kick; Automated moderation too aggressive; False positive moderation actions; No appeal process for bot actions; Moderation logs not accessible to users \\

\bottomrule
\end{tabular}
\end{table*}

\clearpage

\appendices
\section*{Artifact Appendix}
\label{appendix:artifact}
\setcounter{subsection}{0}

The artifact accompanying this paper comprises the analysis pipeline, the
31-theme security and privacy taxonomy, the derived datasets underlying every
reported result, and the BotLens auditing server presented in Section~V. The five
experiments below reproduce the dataset statistics, the thematic and
affective-expression distributions, the permission and metadata analyses, the
inter-rater reliability scores, and both BotLens use cases. Each runs on a
commodity CPU, and the full sequence completes in under twenty minutes, most of
which is package installation.

\subsection{Description \& Requirements}

\subsubsection{How to access}
The artifact is permanently archived on Zenodo at
\url{https://doi.org/10.5281/zenodo.23583221}, and is also available, with any
future updates, at \url{https://github.com/ua-spectral/BotLens}. The data is
included as \texttt{botlens-\allowbreak ndss-\allowbreak data.zip}; unpacking it in the
repository root places \texttt{data/} beside \texttt{config/}, after which no
further assembly is necessary. The code is released under the MIT License and the
data under CC BY-NC-SA 4.0.
\texttt{data/SHA256SUMS.txt} verifies the data component, and
\texttt{ARTIFACT-EVALUATION.md} at the root relates each claim below to the command
that reproduces it.

\subsubsection{Hardware dependencies}
A commodity x86-64 desktop or laptop suffices: eight cores and 16\,GB of memory
are adequate for every step described here, and no GPU is required. Approximately
1\,GB of working space is needed beyond the extracted archive. Regenerating the
dataset from the raw crawl (pipeline Stages 0--4) would require an NVIDIA GPU with
at least 16\,GB of VRAM; because the artifact distributes the outputs of those
stages, evaluators need not repeat them.

\subsubsection{Software dependencies}
Python 3.11 under Linux, macOS or Windows. Every dependency is free software and
is pinned in \texttt{requirements-lock.txt}, a \texttt{pip freeze} of the machine
on which the reported results were produced (Python 3.11.9, \texttt{numpy} 2.3.3,
\texttt{transformers} 4.57.6); dependency resolution has been verified against
PyPI. Ollama and CUDA are required only for the out-of-scope stages noted above.

\subsubsection{Benchmarks}
The data component comprises the 183{,}142 classified reviews, the 8{,}832
security- and privacy-relevant subset annotated with theme and
affective-expression labels, public listing metadata for 23{,}102 bots, the two
annotators' independent labelling workbooks, and a reference rendering of every
generated figure for direct comparison.

\subsection{Artifact Installation \& Configuration}

\begin{verbatim}
python -m venv .venv
source .venv/bin/activate
pip install -r requirements-lock.txt
python -m zipfile -e botlens-ndss-data.zip .
cd data
sha256sum -c SHA256SUMS.txt
cd ..
python -m pytest tests/test_smoke.py -v
\end{verbatim}

\noindent The final command constitutes the minimal working example for the
kick-the-tires stage: it requires no data, no GPU and no network access, and
completes within seconds. All 46 checksums and 17 tests are expected to pass,
which establishes that the artifact is intact and correctly installed.

Experiment (E5) runs the BotLens server through the bundled MCP client,
\texttt{mcp\_server/\allowbreak example\_\allowbreak client.py}, which needs no further
configuration. Registering the server with an MCP application such as Claude
Desktop is optional and described step by step in Section~4 of
\texttt{ARTIFACT-EVALUATION.md}.

\subsection{Major Claims}

\begin{itemize}
\item (C1): The corpus comprises 183{,}142 reviews spanning 23{,}102 bots and 49
detected languages, of which 8{,}832 are security- and privacy-relevant, an
observed pipeline yield of 4.82\%; only 0.77\% of bots reference a privacy policy
in their listings. This is proven by the experiment (E1), reported in Table~I and
Sections~IV-A and IV-D6.
\item (C2): Security and privacy concerns concentrate in Data Practices (71.3\%),
led by Data Sharing (30.5\%) and Data Collection (16.2\%), while confession and
anonymity tags exceed 50\% SPRR. This is proven by the experiment (E2), reported
in Figures~4--6 and Section~IV-B.
\item (C3): Requested permissions and administrator status are weak predictors of
where security and privacy discussion concentrates. This is proven by the
experiment (E3), reported in Figures~10--14 and Section~IV-D.
\item (C4): The teacher-generated labels achieve substantial agreement with two
independent annotators. This is proven by the experiment (E4), reported in
Section~III-C3.
\item (C5): BotLens returns review-derived security and privacy signals as indexed
lookups in the low-millisecond range. This is proven by the experiment (E5),
reported in Sections~V-C and V-D.
\end{itemize}

\subsection{Evaluation}

\subsubsection{Experiment (E1)}
Dataset statistics [2 human-minutes + 1 compute-minute]: establishes the size of
the corpus, the observed pipeline yield, and the rate at which bots disclose a
privacy policy.

\textit{[Preparation]} None beyond installation.

\textit{[Execution]} Run \texttt{scripts/\allowbreak stage7\_\allowbreak analysis.py}, followed by the
metadata command given in Section~3.1 of \texttt{ARTIFACT-EVALUATION.md}.

\textit{[Results]} \texttt{data/\allowbreak results/\allowbreak analysis\_\allowbreak summary.json} reports
\texttt{total\_reviews} 183{,}142, \texttt{privacy\_reviews} 8{,}832 and
\texttt{sprr\_overall} 4.82. The metadata command reports 23{,}102 unique bots, of
which 177 (0.77\%) reference a privacy policy. These are exact counts and no
variation is expected.

\subsubsection{Experiment (E2)}
Themes and tags [2 human-minutes + 1 compute-minute]: regenerates the thematic and
affective-expression distributions underlying RQ1 and RQ2.

\textit{[Preparation]} None.

\textit{[Execution]} Run \texttt{scripts/\allowbreak stage8\_\allowbreak visualizations.py}.

\textit{[Results]} Figures~2--9 are written to \texttt{figures/}, each file named
after its figure number (e.g.\ \texttt{fig02\_sprr\_over\_time.png}).
Theme-group shares are 71.3, 12.8, 9.6 and 3.4\%; theme prevalence rates 30.5,
16.2, 13.3 and 11.1\%; and the highest tag SPRRs 55.4, 55.1, 53.8 and 53.0\%. The
values are deterministic and the figures may be compared directly against the
renderings in \texttt{data/reference\_figures/}. One documented departure from the
published figures is described under Notes.

\subsubsection{Experiment (E3)}
Permissions and metadata [2 human-minutes + 1 compute-minute]: reproduces the
association tests between bot metadata and SPRR reported for RQ3.

\textit{[Preparation]} None.

\textit{[Execution]} Run \texttt{scripts/\allowbreak rq3\_\allowbreak permissions\_\allowbreak metadata.py}.

\textit{[Results]} Figures~10--14 and 18 are written to \texttt{figures/}, and every
statistic of Section~IV-D is printed next to its figure number. The bot subset is
reported as $n=1{,}457$. Permission count and
SPRR are not linearly associated ($r=-0.031$, $p=0.320$); administrator and
non-administrator bots differ negligibly, at 5.69\% ($n=425$) against 5.64\%
($n=1{,}032$); and server count shows the strongest association at Spearman
$\rho=-0.34$. All values are deterministic.

\subsubsection{Experiment (E4)}
Inter-rater reliability [3 human-minutes + 1 compute-minute]: recomputes the
agreement statistics from the two annotators' labelling workbooks.

\textit{[Preparation]} None; the script reads both workbooks from \texttt{data/irr/}.

\textit{[Execution]} Run \texttt{scripts/compute\_kappa.py}.

\textit{[Results]} Cohen's $\kappa$ is 0.7401 between the two annotators, and
0.7136 and 0.8383 between each annotator and the teacher model, computed over
$N=5{,}494$ items with raw agreement exceeding 99.8\% throughout. These correspond
to the values of 0.74, 0.71 and 0.84 reported in Section~III-C3.

\subsubsection{Experiment (E5)}
BotLens [10 human-minutes + 1 compute-minute]: exercises the auditing server and
both use cases presented in Section~V.

\textit{[Preparation]} None. Loading the pre-classified dataset takes approximately
ten seconds and occurs once, when the server starts.

\textit{[Execution]} Run \texttt{mcp\_server/\allowbreak example\_\allowbreak client.py}, which starts
the server over stdio, lists its tools, and runs both use cases; then run
\texttt{mcp\_server/\allowbreak benchmark\_\allowbreak latency.py}, which replays a 22-query session
with the composition reported in Section~V-D.

\textit{[Results]} The server exposes six tools. Wick is reported at 28.9\% SPRR
over 72 of 249 reviews, with a 4.3-star rating, \textsc{administrator} permission,
top themes Data Security, Data Sharing and Data Exposure, and a sentiment
composition of 52.8, 40.3 and 6.9\%, matching Figure~15. The six-bot scan of
Figure~16 reports three bots above 10\% SPRR. Query latency depends on the
evaluating machine; we observe under 1\,ms per call for every tool. The claim under
evaluation is that lookups remain in the low-millisecond range because every
classification is precomputed, rather than the specific timings.

\subsection{Notes}

This artifact corrects a theme-lookup key that did not match the taxonomy's full
spelling for ``Remove Personally Identifiable Information''. One review of 8{,}832
is consequently counted under Data Practices rather than as unmatched, so Figure~4
reads $n=6{,}295$ (71.3\%) and $n=251$ (2.8\%) here, against $n=6{,}294$ and
$n=252$ (2.9\%) in the paper. No other value is affected, and the remaining three
theme groups reproduce exactly. Section~3.2 of \texttt{ARTIFACT-EVALUATION.md}
documents the correction in full.

The active-learning column of Table~III cannot be reproduced from the distributed
artifact, as those runs were not preserved; the LLM-only baselines of 0.727 and
0.938 are present in \texttt{data/\allowbreak results/\allowbreak labeling\_\allowbreak experiment\_\allowbreak results.csv}.
Pipeline Stages 0--6 collectively depend on the raw crawl, a local Ollama instance
and a GPU, and are included as a record of methodology rather than as re-runnable
steps. The scripts generate exactly the paper's Figures~2--14, 17 and 18; Figure~1
is a schematic overview, and Figures~15--16 depict BotLens output rendered in a
report UI, where the presentation is client-dependent although the underlying
values are reproduced exactly by (E5).

Finally, \texttt{scrapers/\allowbreak discord\_\allowbreak server\_\allowbreak scanner.py} authenticates as a Discord
user account, which is contrary to Discord's Terms of Service. Evaluators are
asked not to execute it: no result in the paper depends upon it, and
\texttt{scan\_server\_bots} accepts a plain text file of bot names in its place.

\end{document}